\documentclass[twoside,twocolumn,9pt]{article}
\usepackage{extsizes}
\usepackage[super,sort&compress,comma]{natbib} 
\usepackage[version=3]{mhchem}
\usepackage[left=1.5cm, right=1.5cm, top=1.785cm, bottom=2.0cm]{geometry}
\usepackage{balance}
\usepackage{mathptmx}
\usepackage{sectsty}
\usepackage{graphicx} 
\usepackage{lastpage}
\usepackage[format=plain,justification=justified,singlelinecheck=false,font={stretch=1.125,small,sf},labelfont=bf,labelsep=space]{caption}
\usepackage{float}
\usepackage{fancyhdr}
\usepackage{fnpos}
\usepackage[english]{babel}
\addto{\captionsenglish}{%
  
}
\usepackage{array}
\usepackage{droidsans}
\usepackage{charter}
\usepackage[T1]{fontenc}
\usepackage[usenames,dvipsnames]{xcolor}
\usepackage{setspace}
\usepackage[compact]{titlesec}
\usepackage{hyperref}

\usepackage{epstopdf}

\definecolor{cream}{RGB}{222,217,201}

\begin{document}

\pagestyle{fancy}
\thispagestyle{plain}
\fancypagestyle{plain}{
\renewcommand{\headrulewidth}{0pt}
}

\makeFNbottom
\makeatletter
\renewcommand\LARGE{\@setfontsize\LARGE{15pt}{17}}
\renewcommand\Large{\@setfontsize\Large{12pt}{14}}
\renewcommand\large{\@setfontsize\large{10pt}{12}}
\renewcommand\footnotesize{\@setfontsize\footnotesize{7pt}{10}}
\makeatother

\renewcommand{\thefootnote}{\fnsymbol{footnote}}
\renewcommand\footnoterule{\vspace*{1pt}%
\color{cream}\hrule width 3.5in height 0.4pt \color{black}\vspace*{5pt}} 
\setcounter{secnumdepth}{5}

\makeatletter 
\renewcommand\@biblabel[1]{#1}            
\renewcommand\@makefntext[1]%
{\noindent\makebox[0pt][r]{\@thefnmark\,}#1}
\makeatother 
\renewcommand{\figurename}{\small{Fig.}~}
\sectionfont{\sffamily\Large}
\subsectionfont{\normalsize}
\subsubsectionfont{\bf}
\setstretch{1.125} 
\setlength{\skip\footins}{0.8cm}
\setlength{\footnotesep}{0.25cm}
\setlength{\jot}{10pt}
\titlespacing*{\section}{0pt}{4pt}{4pt}
\titlespacing*{\subsection}{0pt}{15pt}{1pt}

\fancyfoot{}
\fancyfoot[LO,RE]{\vspace{-7.1pt}\includegraphics[height=9pt]{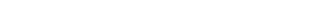}}
\fancyfoot[CO]{\vspace{-7.1pt}\hspace{13.2cm}\includegraphics{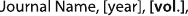}}
\fancyfoot[CE]{\vspace{-7.2pt}\hspace{-14.2cm}\includegraphics{head_foot/RF}}
\fancyfoot[RO]{\footnotesize{\sffamily{1--\pageref{LastPage} ~\textbar  \hspace{2pt}\thepage}}}
\fancyfoot[LE]{\footnotesize{\sffamily{\thepage~\textbar\hspace{3.45cm} 1--\pageref{LastPage}}}}
\fancyhead{}
\renewcommand{\headrulewidth}{0pt} 
\renewcommand{\footrulewidth}{0pt}
\setlength{\arrayrulewidth}{1pt}
\setlength{\columnsep}{6.5mm}
\setlength\bibsep{1pt}

\makeatletter 
\newlength{\figrulesep} 
\setlength{\figrulesep}{0.5\textfloatsep} 

\newcommand{\topfigrule}{\vspace*{-1pt}%
\noindent{\color{cream}\rule[-\figrulesep]{\columnwidth}{1.5pt}} }

\newcommand{\botfigrule}{\vspace*{-2pt}%
\noindent{\color{cream}\rule[\figrulesep]{\columnwidth}{1.5pt}} }

\newcommand{\dblfigrule}{\vspace*{-1pt}%
\noindent{\color{cream}\rule[-\figrulesep]{\textwidth}{1.5pt}} }

\makeatother

\twocolumn[
  \begin{@twocolumnfalse}
{\includegraphics[height=30pt]{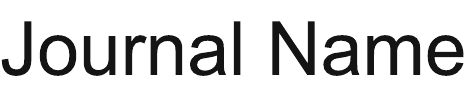}\hfill\raisebox{0pt}[0pt][0pt]{\includegraphics[height=55pt]{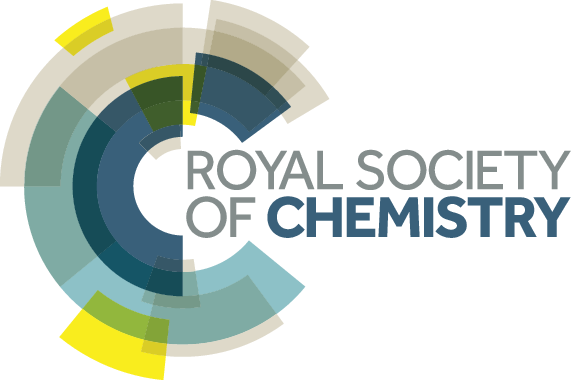}}\\[1ex]
\includegraphics[width=18.5cm]{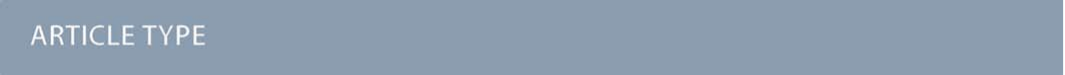}}\par
\vspace{1em}
\sffamily
\begin{tabular}{m{4.5cm} p{13.5cm} }

\includegraphics{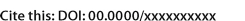} & \noindent\LARGE{\textbf{A Ray Tracing Survey of Asymmetric Operation of the X-ray Rowland Circle Using Spherically Bent Crystal Analyzers}} \\
\vspace{0.3cm} & \vspace{0.3cm} \\

 & \noindent\large{Yeu Chen,$^{\ast}$\textit{$^{a}$} and Gerald T. Seidler,\textit{$^{a}$}} \\

\includegraphics{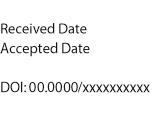} & \noindent\normalsize{
The Spherically Bent Crystal Analyzer (SBCA) is a widely adopted hard x-ray optic, renowned for its good energy resolution and large collection solid angle. It is frequently employed in synchrotron-based techniques like Resonant Inelastic X-ray Scattering (RIXS) and X-ray Emission Spectroscopy (XES), as well as in the rebirth of laboratory-based X-ray Absorption Fine Structure (XAFS) and XES, and its predominant use has been in `symmetric' operation on the Rowland circle. The recent study of Gironda \textit{et al.} (\textit{J. Anal. At. Spectrom.}, 2024, \textbf{39}, 1375) emphasizes the benefits of `asymmetric' SBCA operation, demonstrating the use of multiple crystal reflections from a single SBCA to broaden its accessible energy range. Furthermore, Gironda \textit{et al.} demonstrate that asymmetric operation frequently mitigates energy broadening intrinsic to Johann optics and propose that under a specific Rowland circle configuration, designated here the Johann Normal Alignment (JNA), such broadening is eradicated altogether. We report extensive ray tracing simulations to scrutinize the impact of asymmetric configurations on energy broadening and detector plane defocusing. We find that the performance of asymmetric SBCA operation generally exceeds its symmetric counterpart in energy resolution when no analyzer masking is used and, with strategic detector placement, the decrease in detection efficiency due to defocusing can be minimized. Spectroscopic imaging is adversely affected by the detector plane blurring, but rejection of scattering from special environment windows in x-ray Raman imaging is still feasible. These results help inform future, more common implementation of asymmetric reflections with SBCA.
} \\

\end{tabular}

 \end{@twocolumnfalse} \vspace{0.6cm}

  ]

\renewcommand*\rmdefault{bch}\normalfont\upshape
\rmfamily
\section*{}
\vspace{-1cm}


\footnotetext{\textit{$^{a}$~Department of Physics, University of Washington, Seattle, WA, USA. Email: seidler@uw.edu}}




\section{Introduction}

\; \;

The spherically bent crystal analyzer (SBCA) is one of the most widely used hard x-ray optics due to its favorable combination of fine energy resolution and large collection solid angle \cite{johann31}. SBCAs are used in synchrotron x-ray light source endstations that perform Resonant Inelastic X-ray Scattering (RIXS) \cite{huotari14, moretti18, robledo18}, X-ray Emission Spectroscopy (XES) \cite{kleymenov11, llorens12, sokaras13}, and X-ray Raman Scattering \cite{fister06, sokaras12, huotari17}.  SBCAs are also the most commonly used x-ray optic in the ongoing rebirth of laboratory-based X-ray Absorption Fine Structure (XAFS) and X-ray Emission Spectroscopy (XES) \cite{seidler14, seidler16, mortensen16, mortensen17, bes18, mundy18, jahrman19, zimmermann20}.

\begin{figure}[!h]
\centering
  \includegraphics[width=\textwidth/2-1cm]{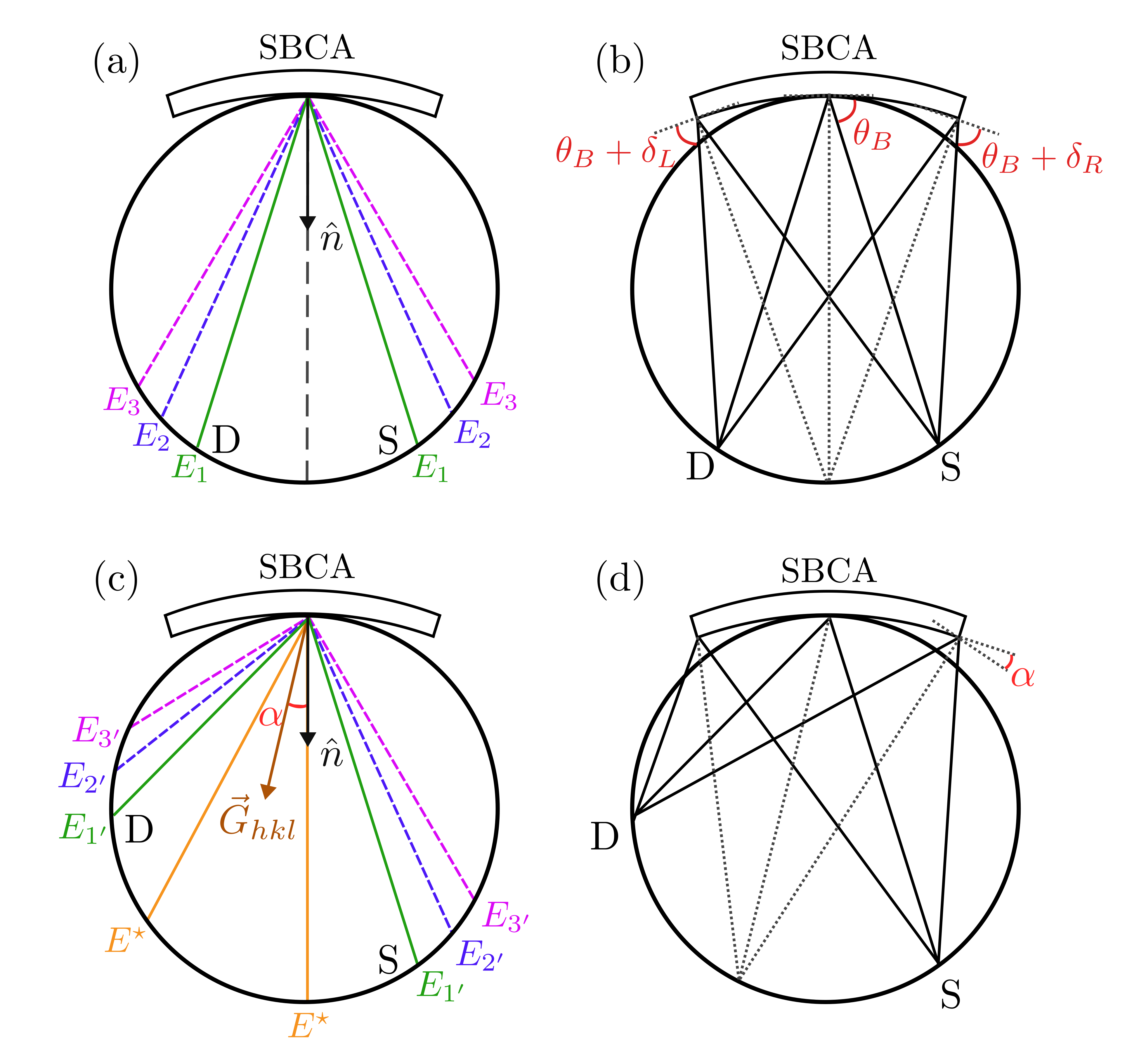}
  \caption{(a) By positioning the source (S) and detector (D) symmetrically on the Rowland circle, the detector selects the energy to be detected through Bragg diffraction. (b) By setting the SBCA bent radius to be twice as the Rowland circle, all diffracted rays from the analyzer are (approximately) focused onto the detector. (c) When operating asymmetrically a new reflection plane $\vec{G}_{hkl}$ is introduced. The $E^{\star}$ lines mark the configuration where $\theta_B + \alpha = 90 \, \text{deg}$. (d) The diffracted rays are still approximately focused onto the detector, with slight blurring which will be discussed later in this paper.}
  \label{symm_vs_asymm}
\end{figure}

This broad and successful implementation of SBCAs has almost exclusively used ``symmetric" operation on the Rowland circle (see Figure \ref{symm_vs_asymm}a and \ref{symm_vs_asymm}b). The exceptions have come from small corrections in wafer miscut for laboratory spectrometers \cite{mortensen17}, rare cases in synchrotron inelastic x-ray scattering when a desired SBCA orientation was not available but could be accessed with modest deviation from the nominal surface planes of an available SBCA \cite{seely19}, and in the recent work of Giranda \textit{et al.} \cite{gironda24}.

In Giranda \textit{et al.} \cite{gironda24}, it was demonstrated that an asymmetric operation of the SBCA on the Rowland circle offers significant advantages that may have been underutilized. In particular, it provides the opportunity to harness a multitude of crystal reflections from a single analyzer, thus widening the energy range accessible using any single SBCA and enabling a natural automation to access a wide energy range without need for exchanging and aligning multiple SBCA. Moreover, when properly configured, asymmetric operation can greatly reduce, and sometimes fully eliminate, energy broadening from so-called Johann error.

The source of the Johann error is due to the diffractive crystal being spherically bent to twice the radius of the Rowland circle, resulting a gap between the crystal periphery and the Rowland circle (Figure \ref{symm_vs_asymm}b). This causes a Bragg angle deviation at the analyzer edges in the Rowland plane and consequent energy broadening. When operating asymmetrically (Figure \ref{symm_vs_asymm}c and \ref{symm_vs_asymm}d), a new angular parameter $\alpha$ is introduced which is defined to be the angle between the SBCA wafer normal and the lattice vector, $\vec{G}_{hkl}$, of the chosen new reflection (or equivalently the angle between the crystal face and the selected reflection plane). It should be noted that the crystal normal vector continues to be oriented toward the bottom of the Rowland circle, and $\vec{G}_{hkl}$ bisects the incident and the reflected rays.  When an additional mechanical degree of freedom, $\phi$, is added to rotate the SBCA about its own cylindrical axis the spectrometer user can then bring any desired $\vec{G}_{hkl}$ into the Rowland plane.  This is the basis of the demonstration in Gironda, et al. of performance over a wide energy range with a single SBCA.

\begin{figure}[!h]
\centering
  \includegraphics[width=\textwidth/2-2cm]{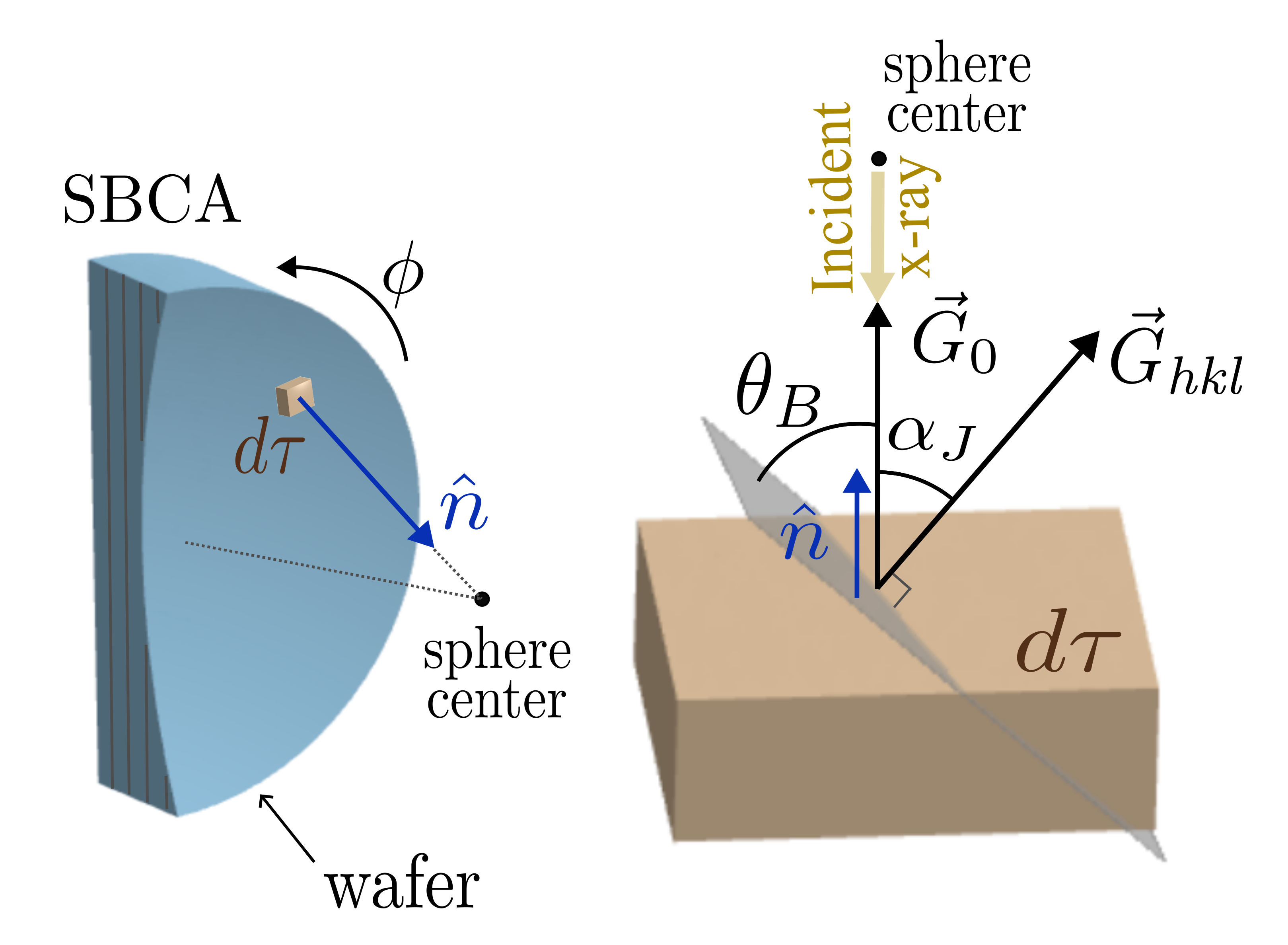}
  \caption{Graphical representation of the Johann normal aligment (JNA). The left figure shows a sectional view of the SBCA and all crystal element ($d\tau$) normal vectors point towards the sphere center of the SBCA. The right figure shows one specific crystal element, where incident x-ray is along the crystal normal and maintains a constant angle $\alpha_J$ with respect to $\vec{G}_{hkl}$.}
  \label{xtal}
\end{figure}

As discussed previously by Suortti \textit{et al.} \cite{suortti99}, when $\theta_B + \alpha = 90 \, \text{deg}$ the source coincides precisely with the sphere center of the SBCA (see $E^{\star}$ lines in Figure \ref{symm_vs_asymm}c and see Figure \ref{xtal}). For a given $\theta_B$, we define such $\alpha$ as $\alpha_J$, i.e. $\alpha_J = 90 \, \text{deg} -\theta_B$. In this conceptually important configuration and ignoring bending strains, the incident X-ray beam impinges normally at all points on the wafer surface with a constant angle $\theta_B$ between the incident ray and the chosen reflection plane (Figure \ref{xtal}). As a result, a Bragg angle of $90 \, \text{deg}- \alpha_J$ is maintained at every point on the spherically-curved wafer face of the optic, effectively eliminating the Johann error. We call this optimal configuration the Johann Normal Alignment, henceforth JNA.

Here, we present an extensive geometric ray tracing study of asymmetric Rowland circle operation of SBCA’s under conditions immediately relevant for experiment. We focus on two key issues.  First, we examine the energy response function of SBCAs as a function of experimental configuration with an emphasis on suppression of Johann error. We find that the Johann error is completely suppressed at JNA, and we provide a broad survey of the benefit of asymmetric operation on energy resolution. Second, we use a purely geometrical calculation (hence ignoring penatration and analyzer strain effects \cite{honkanen14, honkanen20}) to address what may be the largest drawback to asymmetric operation of SBCA, i.e., the poorer focus on the detector and especially the increased sagittal defocusing. We discuss the effects of this defocusing under two cases, namely the decrease in detection efficiency due to limited detector sensor area and the degradation in spatial resolution when doing X-ray Raman Imaging. These results directly inform the case-specific decision of best practice for asymmetric operation constrained by the available detector size, and thus seek to improve future experiment design using SBCA in asymmetric configurations.

\section{Method}

\; \;
To computationally implement the Rowland geometry, we use the open-source ray tracing software XRay Tracer (xrt) \cite{klementiev14}. Though the xrt package has the capability to calculate x-ray diffracted by deformed crystal, we found that the stress and strain effect has minimal influence on our results (contributes less than 0.5 eV of broadening when at symmetric $\theta_B = 60 \, \text{deg}$, and 0.2 eV when under JNA). We expect this effect to be even smaller when analyzers with strain relief cuts are used. Hence, our simulation is purely geometric. A typical run of simulation includes three distinct optical elements: an  x-ray source, a SBCA, and a detector. The source is circular with a diameter of 50 µm, characterized by a flat energy distribution spanning a specified energy range. The number of incident rays ranges from 5M to 50M between studies. The SBCA is a Si(551) analyzer, possessing a bend radius of 500 mm and a crystal face diameter of 100 mm; these sizes match SBCA's most often used in synchrotron and laboratory applications. The detector has a circular face, featuring an active area of 150 mm$^2$; this matches the dimensions of the widely employed KETEK AXAS-M detector. The spatial placement of these components on the Rowland circle (with a radius of 250 mm) is determined by the prescribed $\theta_B$ and $\alpha$.

\begin{figure*}[!]
 \centering
 \includegraphics[width=\textwidth-2cm]{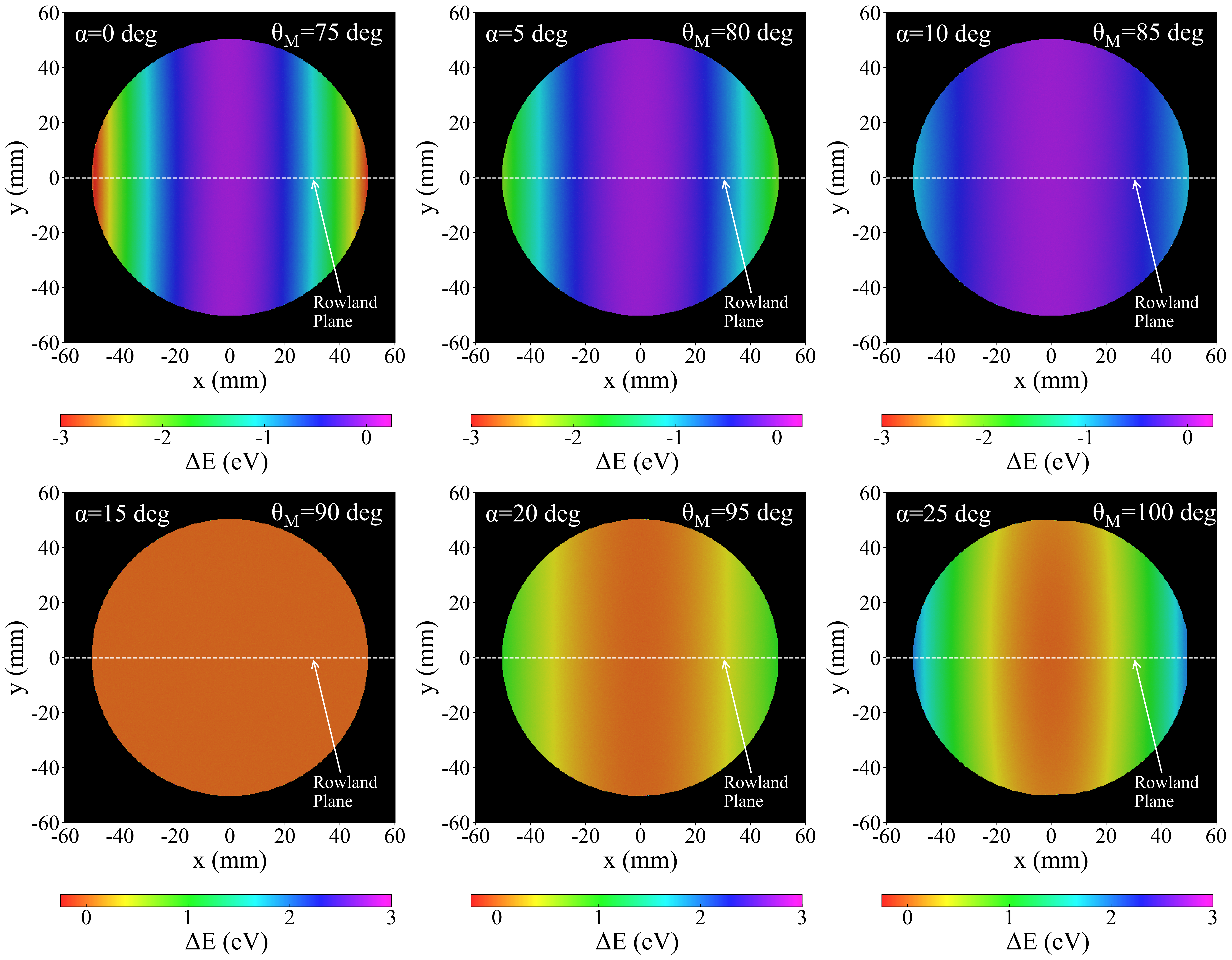}
 \caption{Crystal analyzer face images of a Si(551) analyzer operating at $\theta_B = 75 \, \text{deg}$ (corresponding to $E_0 = 8439 \, \text{eV}$) and multiple values of the asymmetry $\alpha$. The mechanical angle of the spectrometer is indicated by $\theta_M$, which is defined through the relation $\theta_M = \theta_B + \alpha$. The simulation employs a source size of 50 $\mu$m. At $\alpha=\alpha_J=15 \, \text{deg}$ the Johann error is eliminated.}
 \label{study1_xtal_face}
\end{figure*}

\section{Results and Discussion}

\subsection{Johann Error Characterization}
\; \;
In Figure \ref{study1_xtal_face}, a sequence of crystal analyzer face images is depicted at various $\alpha$ for $\theta_B=75 \, \text{deg}$. The coloration in these images signifies the energy deviation from the energy $E_0$ due to Johann error. Under symmetric operation ($\alpha=0 \, \text{deg}$), the energy deviation reaches -3 eV at the crystal edge. Conversely, at $\alpha=\alpha_J=15 \, \text{deg}$ (where $\alpha_J$ is the required angle for achieving JNA for a given $\theta_B$), the energy deviation approaches zero and displays a uniform distribution. As $\alpha$ surpasses $\alpha_J$, Johann error re-emerges, but manifesting a positive energy deviation.

\begin{figure}[!h]
\centering
  \includegraphics[width=\textwidth/2-2cm]{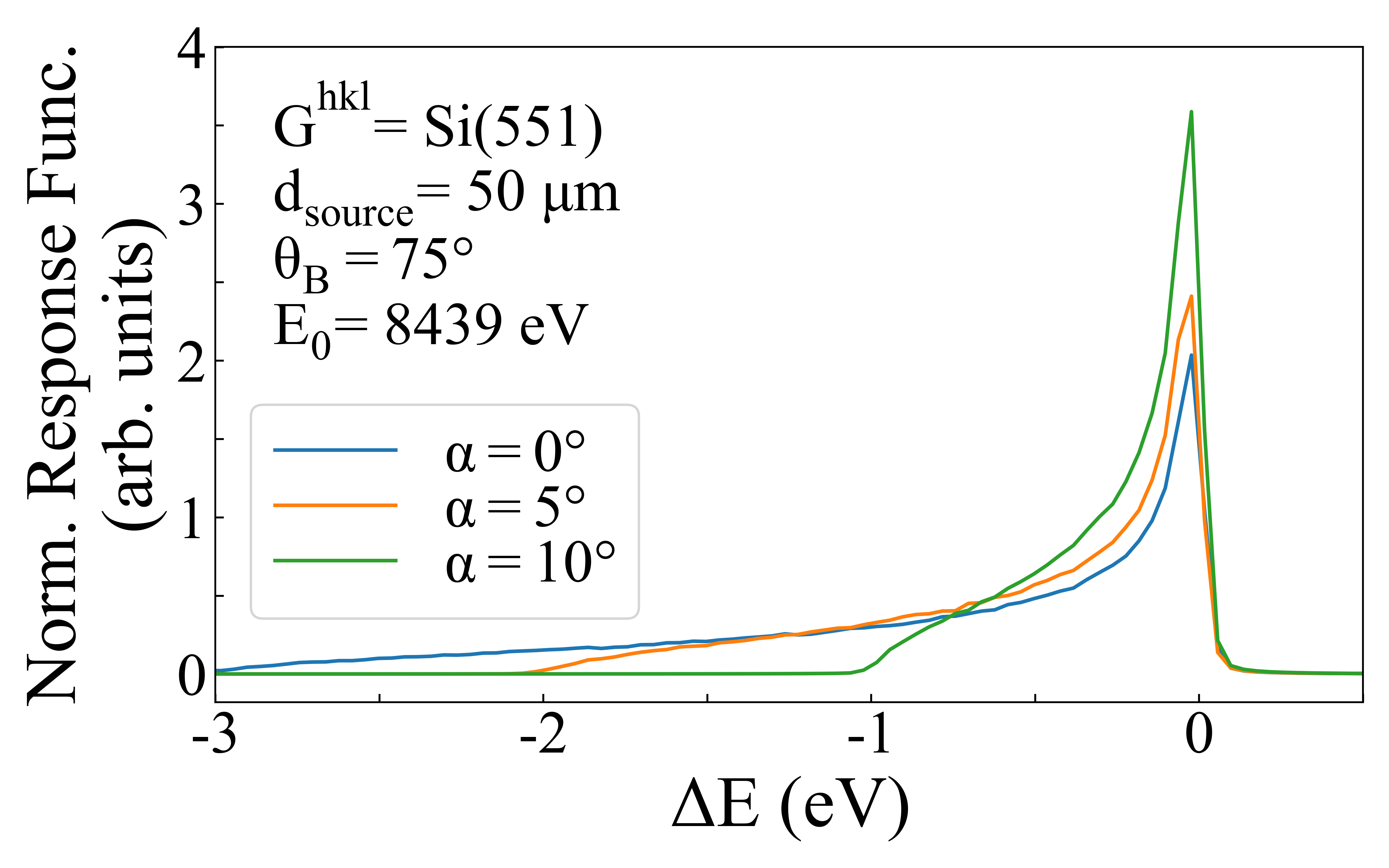}
  \includegraphics[width=\textwidth/2-2cm]{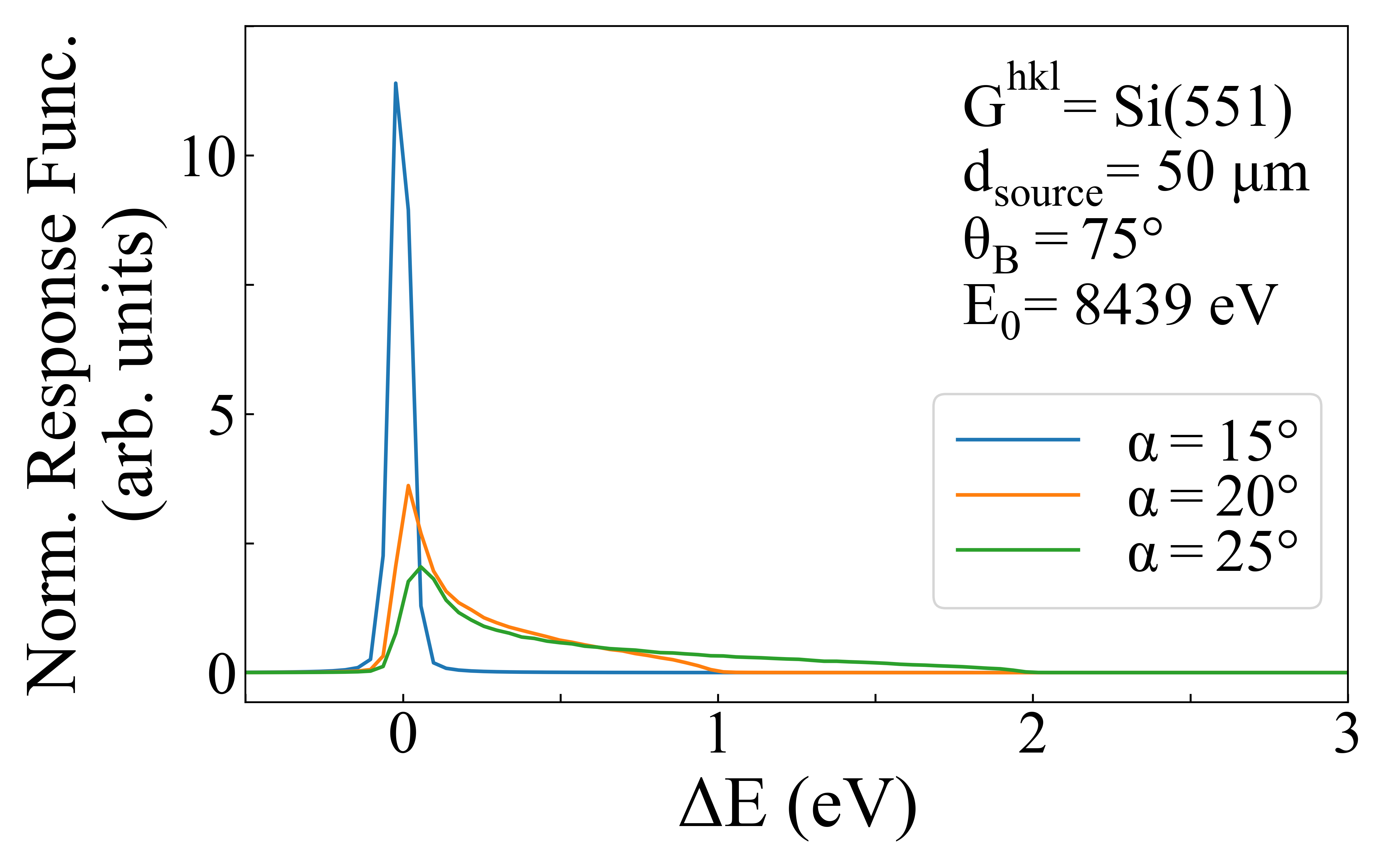}
  \caption{Integral normalized response functions corresponding to each crystal face images shown in Figure \ref{study1_xtal_face}.}
  \label{study1_RF}
\end{figure}

In Figure \ref{study1_RF} we show the energy response functions corresponding to the analyzer face images presented in Figure \ref{study1_xtal_face}. At $\alpha = \alpha_J = 15 \, \text{deg}$ the response function is symmetrical about 0 eV with a FWHM of approximately 0.04 eV attributed to source size broadening. For $\alpha<\alpha_J$ the response functions display a low-energy tail, whereas for $\alpha>\alpha_J$ they instead exhibit a high-energy tail.

\begin{figure}[!h]
\centering
  \includegraphics[width=\textwidth/2-2cm]{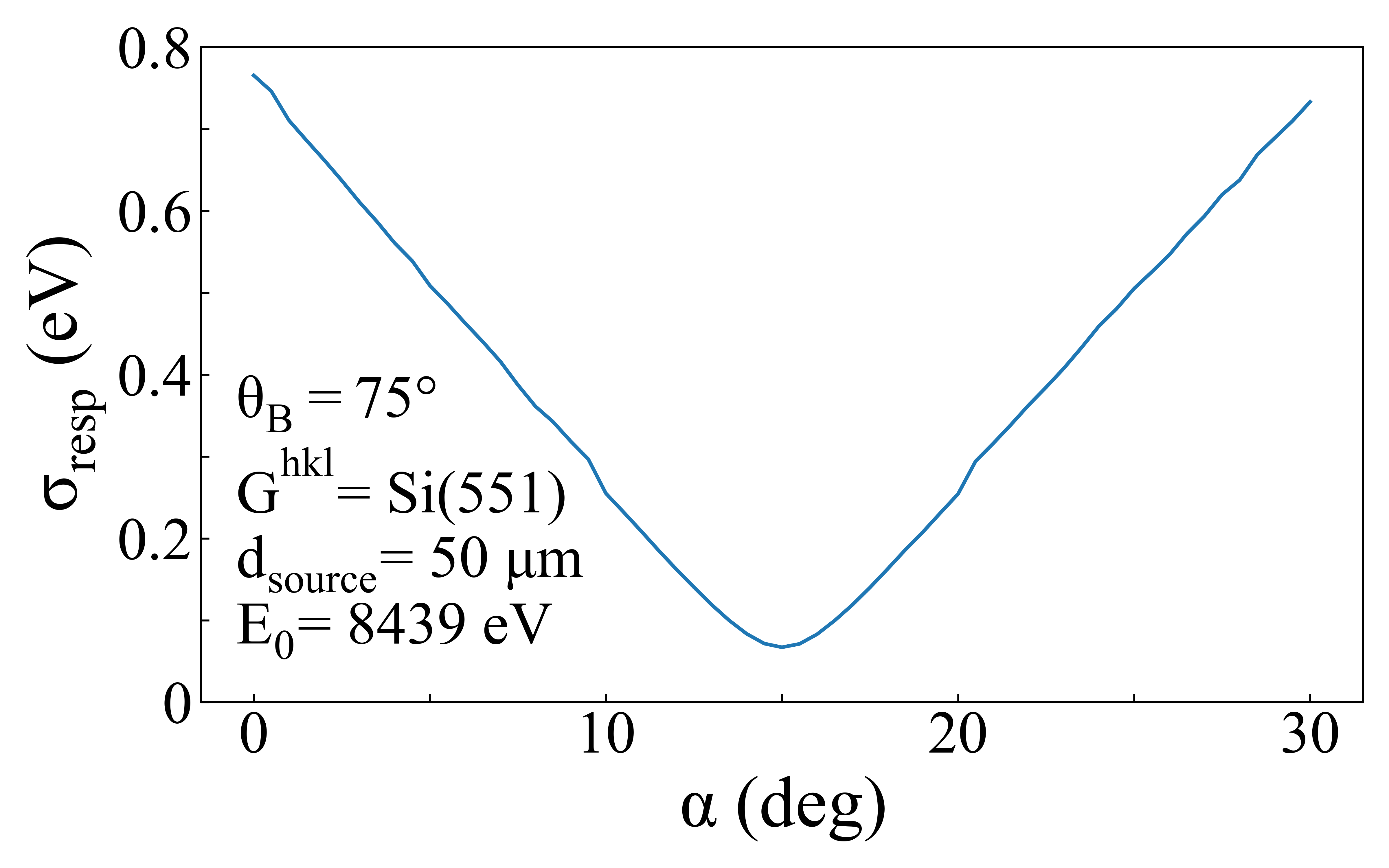}
  \caption{Standard deviation of the response functions as a function of $\alpha$, when $\theta_B$ is 75 deg. A Si(551) analyzer is used which gives $E_0 = 8439 \, \text{eV}$.}
  \label{study1_FOM}
\end{figure}

In Figure \ref{study1_FOM} we show the standard deviation of the response function ($\sigma_{resp}$) as a function of $\alpha$. The standard deviation decreases from 0.77 eV when $\alpha=0 \, \text{deg}$ (symmetric operation) to 0.12 eV at JNA when $\alpha = \alpha_J= 90 \, \text{deg} -\theta_B$. Recall that the broadening for $\alpha = \alpha_J$ is due to the finite source size.

\begin{figure}[!h]
\centering
  \includegraphics[width=\textwidth/2-2cm]{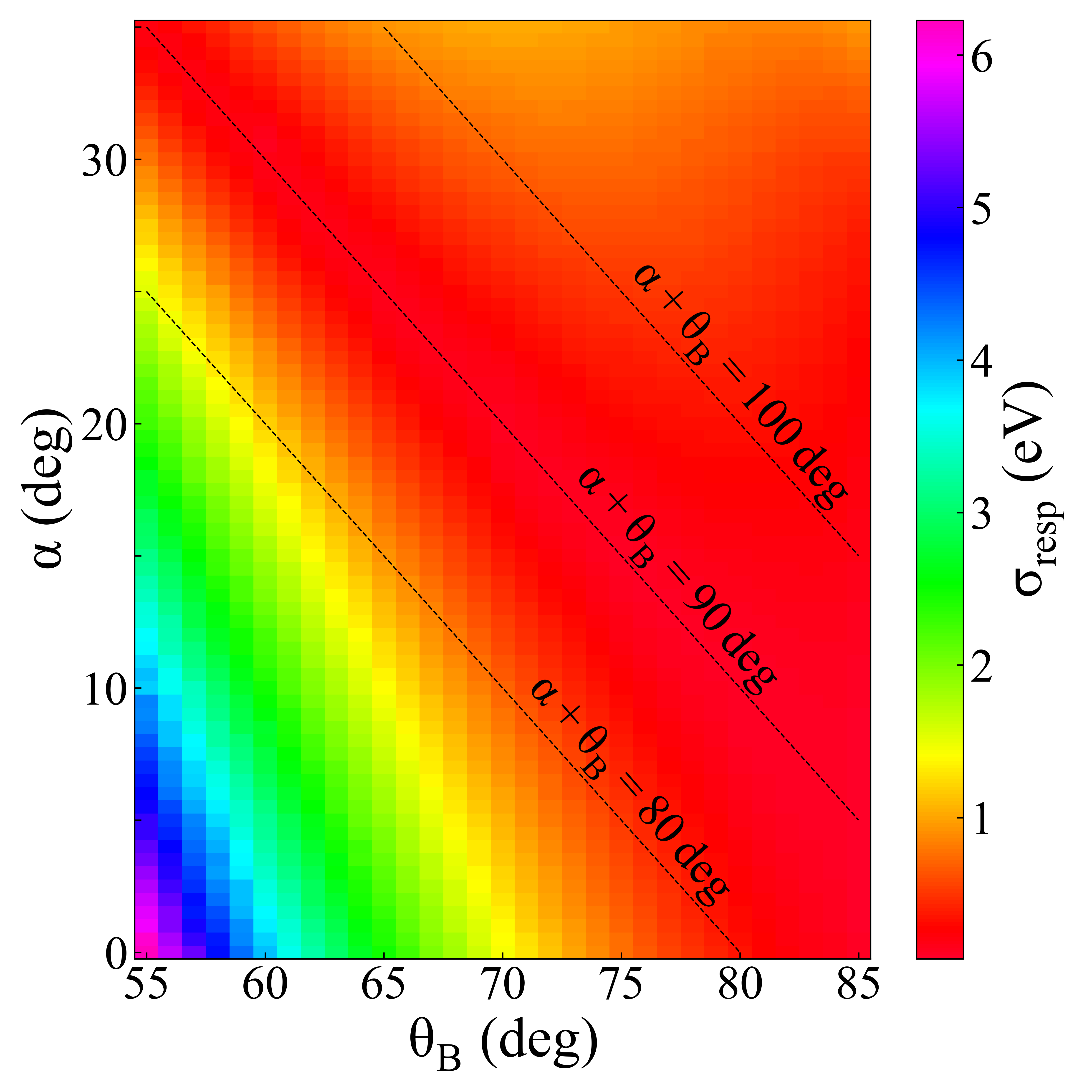}
  \caption{Standard deviation of the response function ($\sigma_{resp}$) as a function of $\alpha$ and $\theta_B$, presented as a 2-D desity map.}
  \label{std_contour}
\end{figure}

In Figure \ref{std_contour} we provide a more complete exploration of energy broadening in the 2-D space of $\theta_B$ and $\alpha$. Note that $\sigma_{resp}$ exhibits a minimum along the dashed line where $\alpha + \theta_B = 90 \, \text{deg}$. Also note that $\sigma_{resp}$ remains small, i.e. around the range of an eV, within the range of $\alpha + \theta_B > 80 \, \text{deg}$, . However, $\sigma_{resp}$ increases rapidly when $\alpha+\theta_B$ drops below 80 degrees. With this we can provide a recommendation range of $\theta_B$ and $\alpha$ when running an experiment with an unmasked analyzer to maximize solid angle

\begin{figure}[!h]
\centering
  \includegraphics[width=\textwidth/2-2cm]{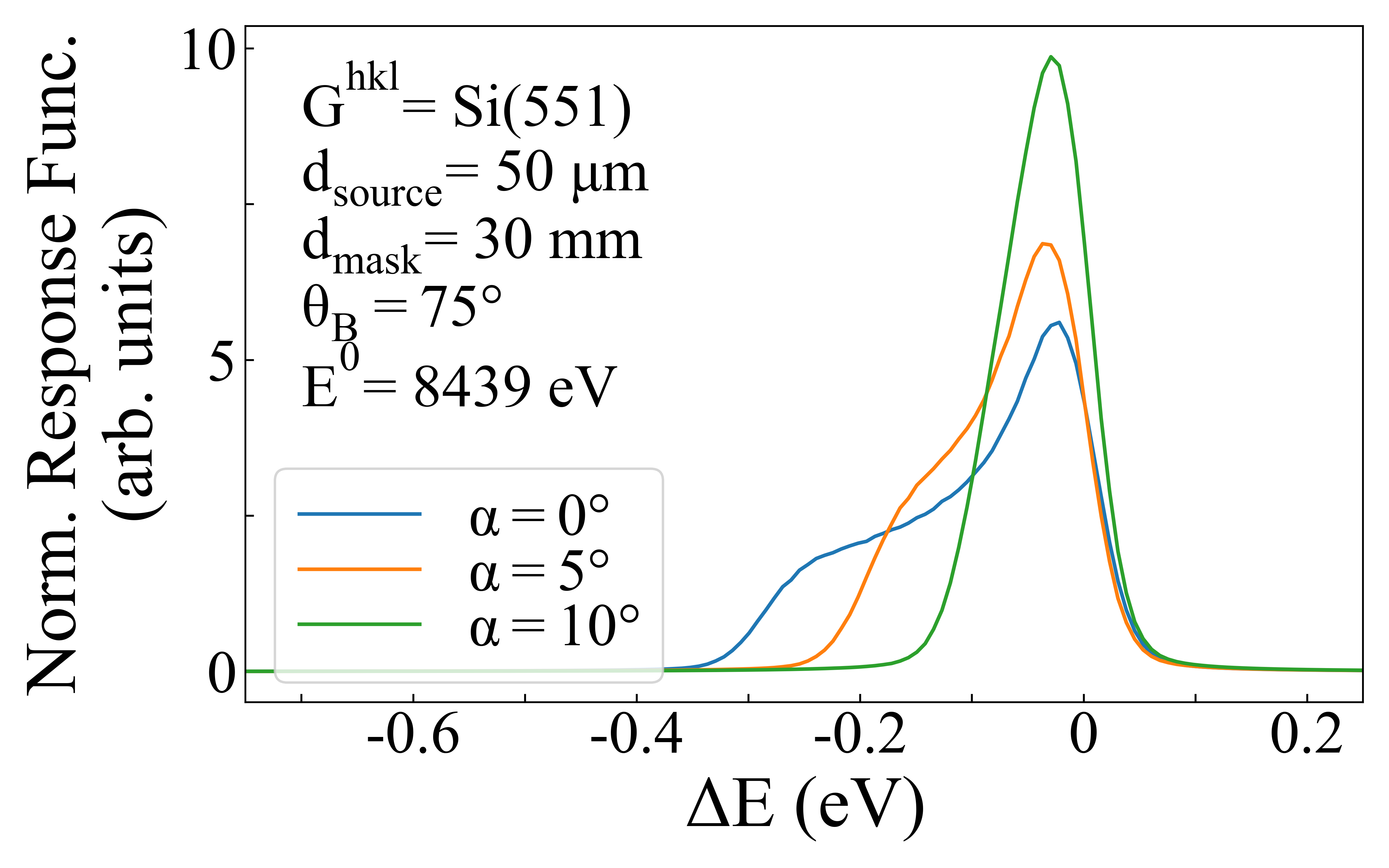}
  \includegraphics[width=\textwidth/2-2cm]{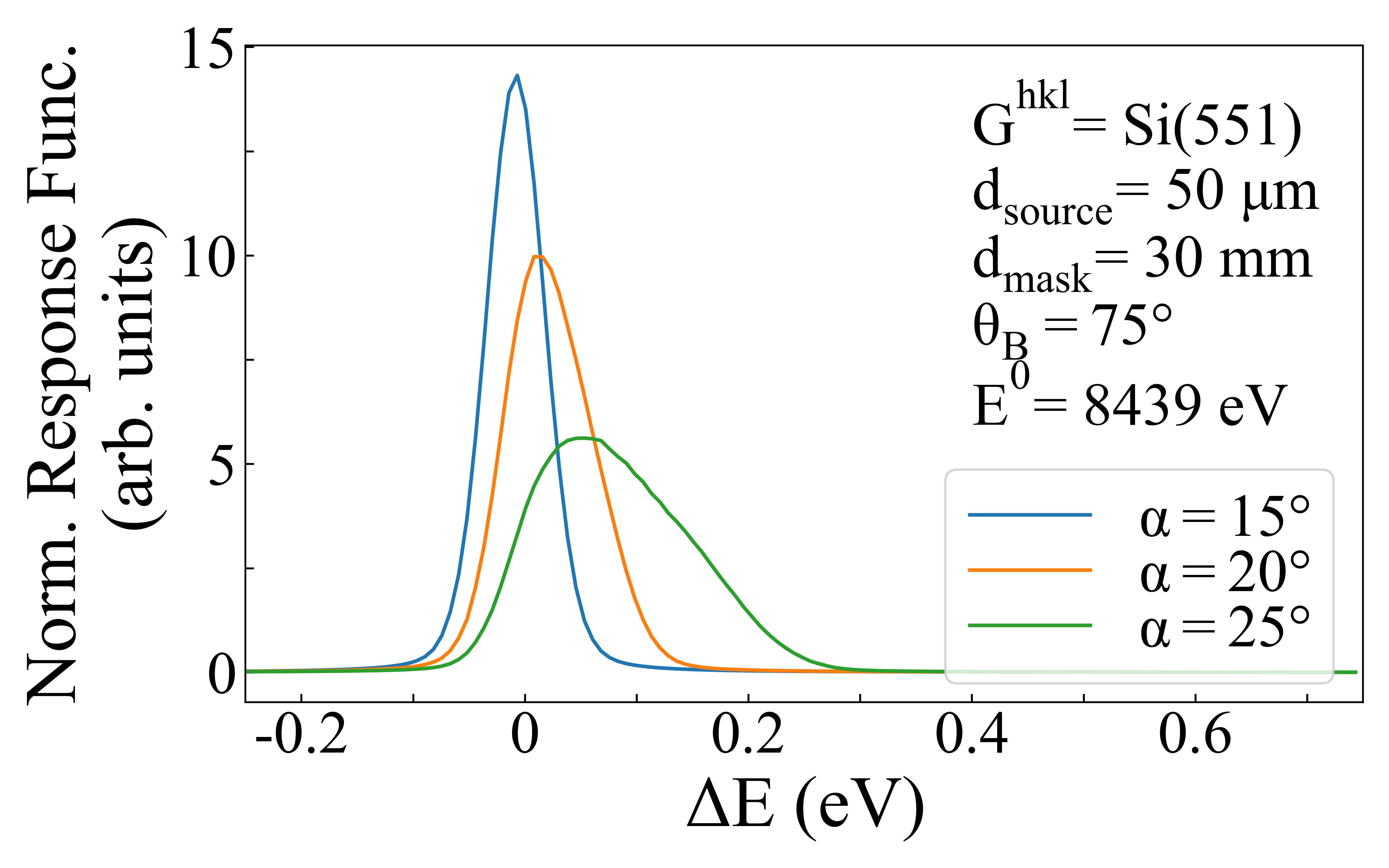}
  \caption{Integral normalized response functions at $\theta_B = 75 \, \text{deg}$ and selected values for $\alpha$. The simulation parameters are the same as in Figure \ref{study1_RF} but with a 30-mm wide analyzer mask.}
  \label{study3_RF}
\end{figure}

On the other hand, if an experiment requires $\alpha + \theta_B < 80 \, \text{deg}$ then masking the analyzer edges should be considered, just as is commonly done for symmetric operation. Hence we repeat the simulation of Figure \ref{study1_RF} with a 30-mm wide analyzer mask (see Figure \ref{study3_RF}), which gives a 38$\%$ of active area comparing to the unmasked case. Unsurprisingly, we observe the response function standard deviations are greatly reduced for all $\alpha$ compare to Figure \ref{study1_RF}, except when $\alpha = \alpha_J = 15 \,\text{deg}$ where the response function is unchanged.

In conclusion, through examining the response functions under various cases, we see that the energy resolution is greatly improved through asymmetric operation, especially when $\alpha$ is close to $\alpha_J$. Applying a mask to the SBCA edges in the Rowland plane, as often done in symmetric case, further reduces the energy broadening. This facilitates the use of multiple reflection planes within a single SBCA without the concern of compromising energy resolution.

\subsection{Detection Efficiency Study} \label{det eff}
\begin{figure*}[!]
 \centering
 \includegraphics[width=\textwidth-4cm]{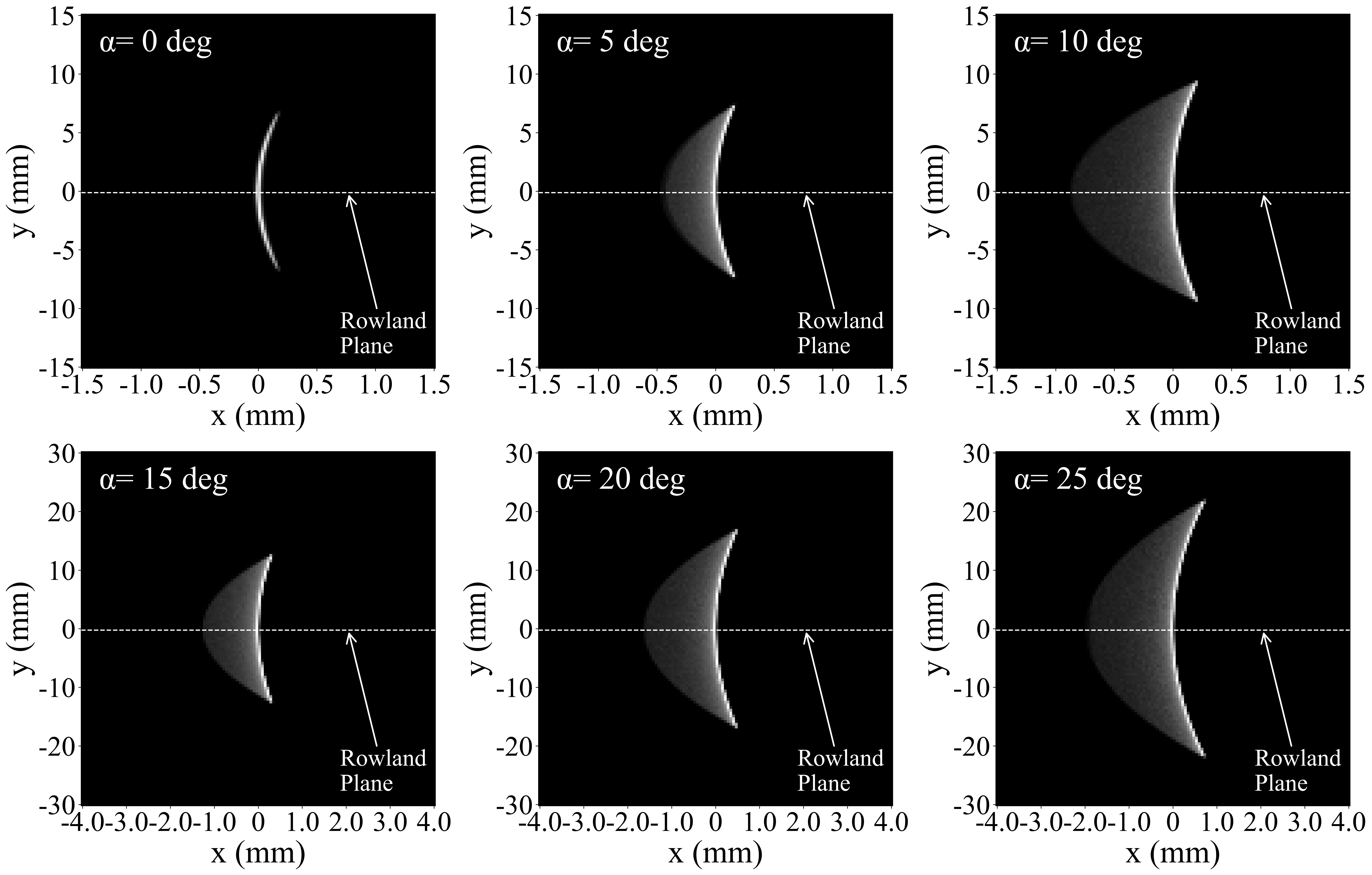}
 \caption{Detector plane images at the indicated values of $\alpha$ when $\theta_B = 75 \, \text{deg}$. A noticeable defocusing in the vertical direction can result in decreased detection efficiency because of finite detector size.}
 \label{study_4_det_imgs}
\end{figure*}

\; \; 
While SBCA-based spectrometers can use many different detectors, the strong background rejection of silicon drift detector (SDD) makes them the most popular choice. In the context of asymmetric operation, the typical SDD active diameter ($D_{SDD}=$ 13.8 mm) raises the issue of signal losses. The key issue is shown in Figure \ref{study_4_det_imgs} where intensity maps are shown for the detector plane at the Rowland circle at $\theta_B = 75 \, \text{deg}$ and various $\alpha$. As $\alpha$ increases, both the width and especially the height of the detected beam increase. Fortunately, as we shown here, this situation can be improved by strategically placing the detector behind the Rowland circle. To explore how detection efficiency can be optimized through detector placement, we first present a geometrical explanation on how beam shape changes from the point behind the Rowland circle.

\begin{figure*}[!]
 \centering
 \includegraphics[width=\textwidth-4cm]{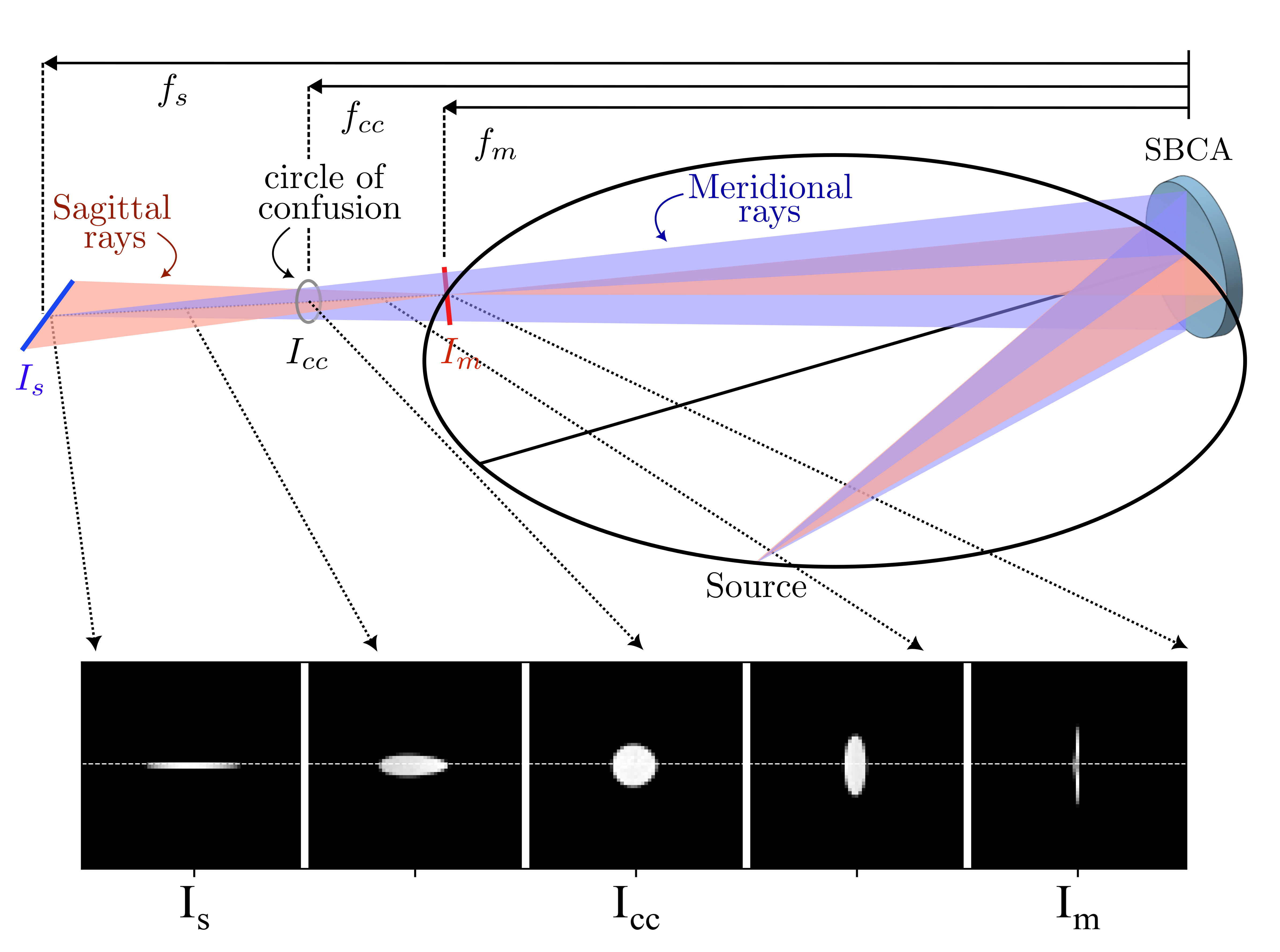}
 \caption{Astigmatic imaging errors of an SBCA result in a vertical line focus ($I_m$) from the meridional rays on Rowland circle and a horizontal line focus ($I_s$) from the sagittal focus behind the circle. The position of the circle of confusion ($I_{cc}$) is where the beam height equals the beam width. The detector images at $I_m$, $I_{cc}$, and $I_s$ (with $\theta_B=75 \, \text{deg}$ and $\alpha=0 \, \text{deg}$) are shown at the bottom, with additional images at the midpoints of each segment.}
 \label{vert_vs_horiz_focus}
\end{figure*}

While Johann crystal analyzers offer good on-circle focusing in the Rowland plane when employing a point source, as illustrated in Figure \ref{symm_vs_asymm} (b) and (d), the vertical focusing exhibits a finite height, see Figure \ref{vert_vs_horiz_focus}. This is because the SBCA focal point in the meridional plane is behind the Rowland circle, consequently resulting in a vertical (meridional) line focus of the point source on the Rowland circle and a horizontal (sagittal) line focus positioned behind it.

The distance from the vertical focus ($I_m$) to the crystal is represented as $f_m$, whereas the corresponding distance for the horizontal focus ($I_s$) is labeled as $f_s$. Under symmetrical configuration, these distances have been well studied \cite{bitter08}, and similar calculations for the asymmetric case give
\begin{equation}
    f_m = R \, \sin (\theta_B - \alpha)
\end{equation}
and
\begin{equation}
    f_s =-\frac{R \, \sin^2(\theta_B + \alpha)}{\sin(\theta_B - \alpha) \, \cos(2(\theta_B+\alpha))}
    \label{eq: fs}
\end{equation}
where R is the radius of curviature. The beam height at the vertical focus ($h_m$) and the width at the horizontal focus ($w_s$) are
\begin{equation}
    h_m = 2 (f_s-f_m) \tan(\frac{d_{xtal}}{f_s})
    \label{fv}
\end{equation}
and
\begin{equation}
    w_s = 2 (f_s-f_m) \tan (\frac{\gamma}{2})
\end{equation}
where $\gamma$ is the sagittal angular size of the analyzer viewed from the source and $d_{xtal}$ is the analyzer diameter.

Using these expressions, we then can define the distance $f_{cc}$ from the crystal face to the minimum circle of confusion (CC) at location $I_{cc}$, i.e. where the beam height equals the beam width
\begin{equation}
    f_{cc} = f_m + \frac{h_s(f_s-f_m)}{h_m +w_s} \; \text{.}
\end{equation}
The analytical results inform the ray tracing needed for a fully quantitative picture.

\begin{figure*}[!]
 \centering
 \includegraphics[width=\textwidth-4cm]{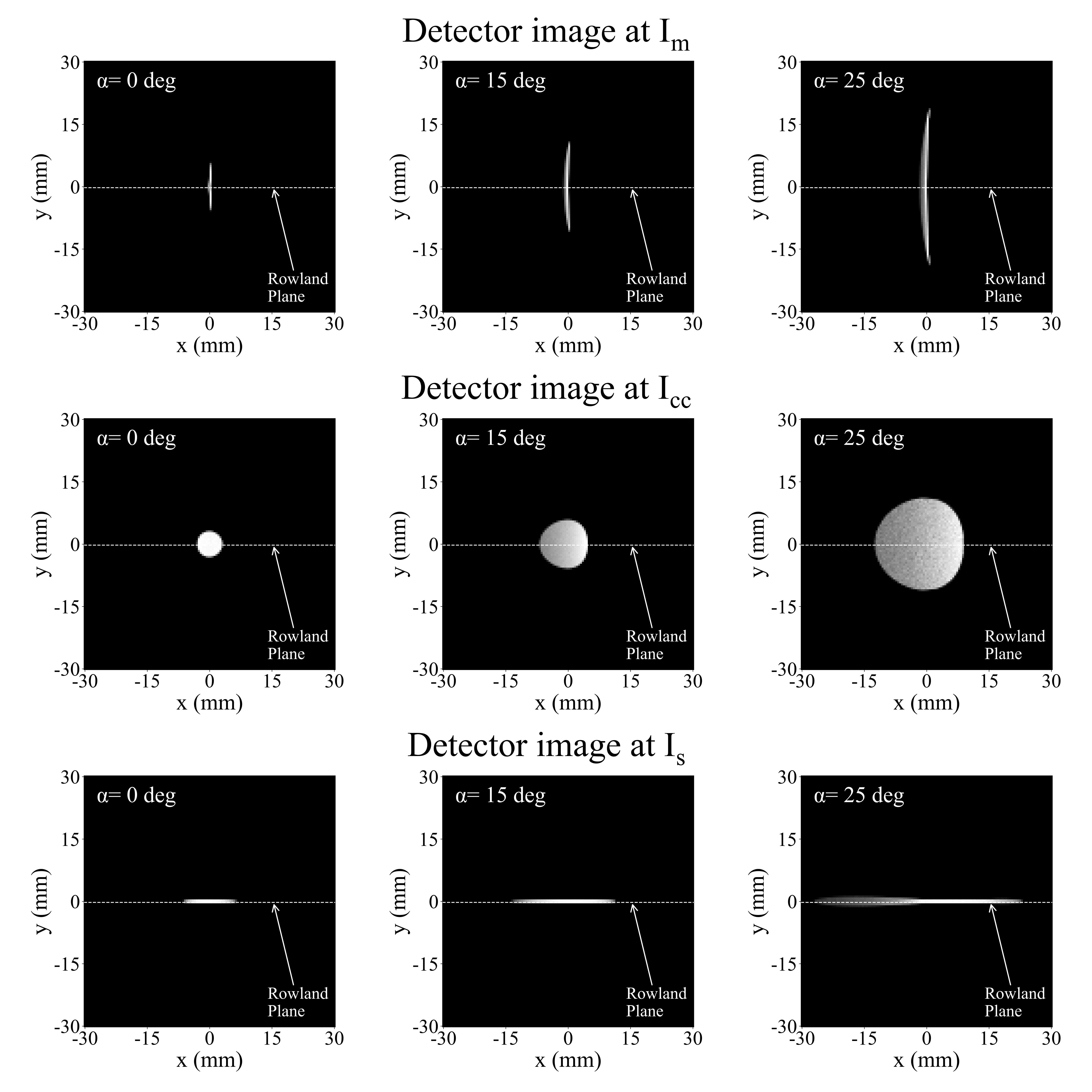}
 \caption{Detector plane images at $\theta_B = 75 \, \text{deg}$ and 
$\alpha=$0, 15, and 25 deg. Each row from top to bottom correspond to $I_m$, $I_{cc}$, and $I_s$ respectively.}
 \label{det_pos_scan_img_75}
\end{figure*}

In Figure \ref{det_pos_scan_img_75} we extend the series of detector images presented in Figure \ref{vert_vs_horiz_focus} to the asymmetric case, where $\alpha =15 \, \text{deg}$ (middle column) and $\alpha=25 \, \text{deg}$ (right most column). Notice in all cases, the beam spot is most compatible with the circular SDD active region near $I_{cc}$, suggesting that repositioning the detector strategically could improve detection efficiency.

\begin{figure}[!h]
\centering
  \includegraphics[width=\textwidth/2-4cm]{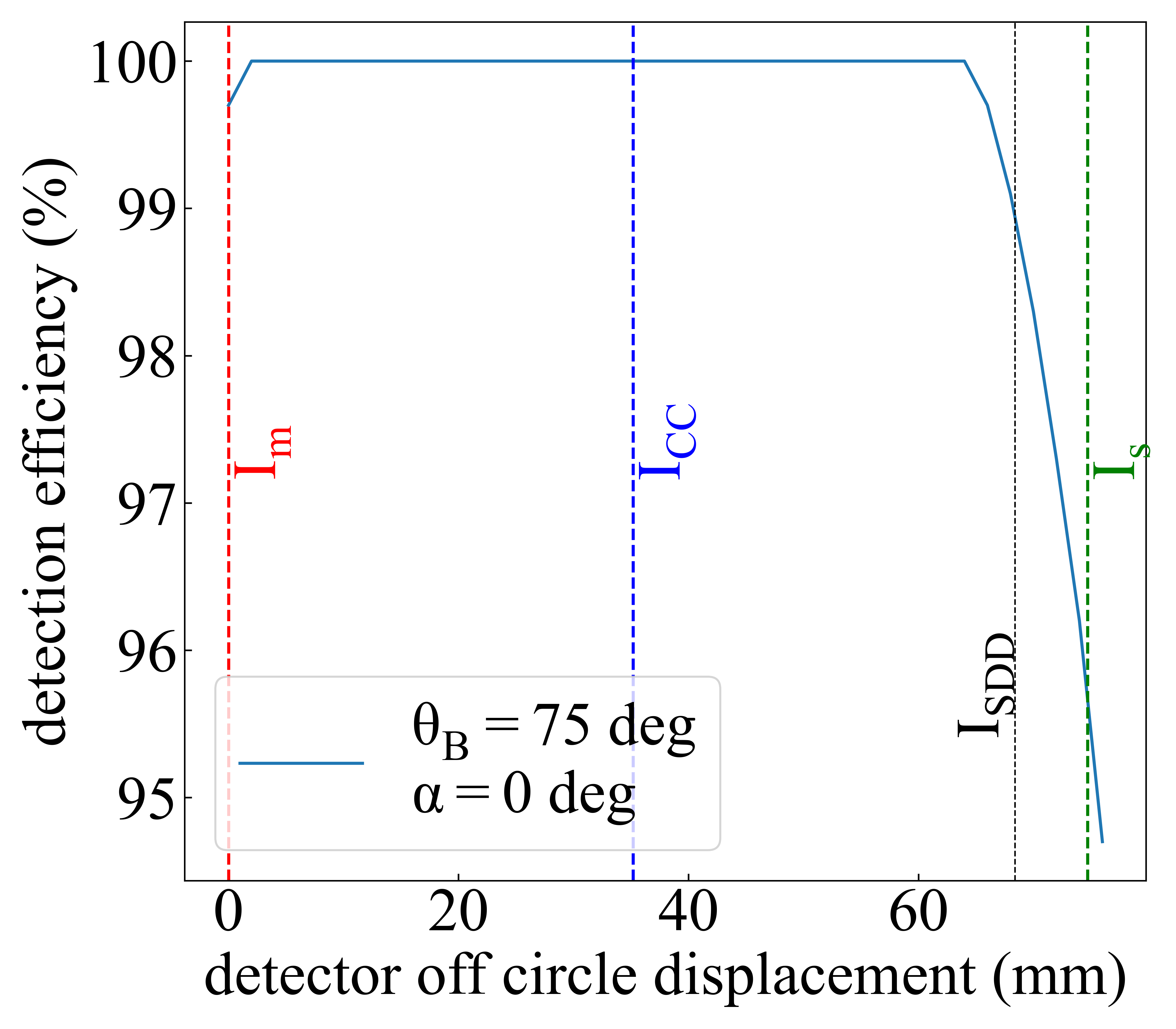}
  \includegraphics[width=\textwidth/2-4cm]{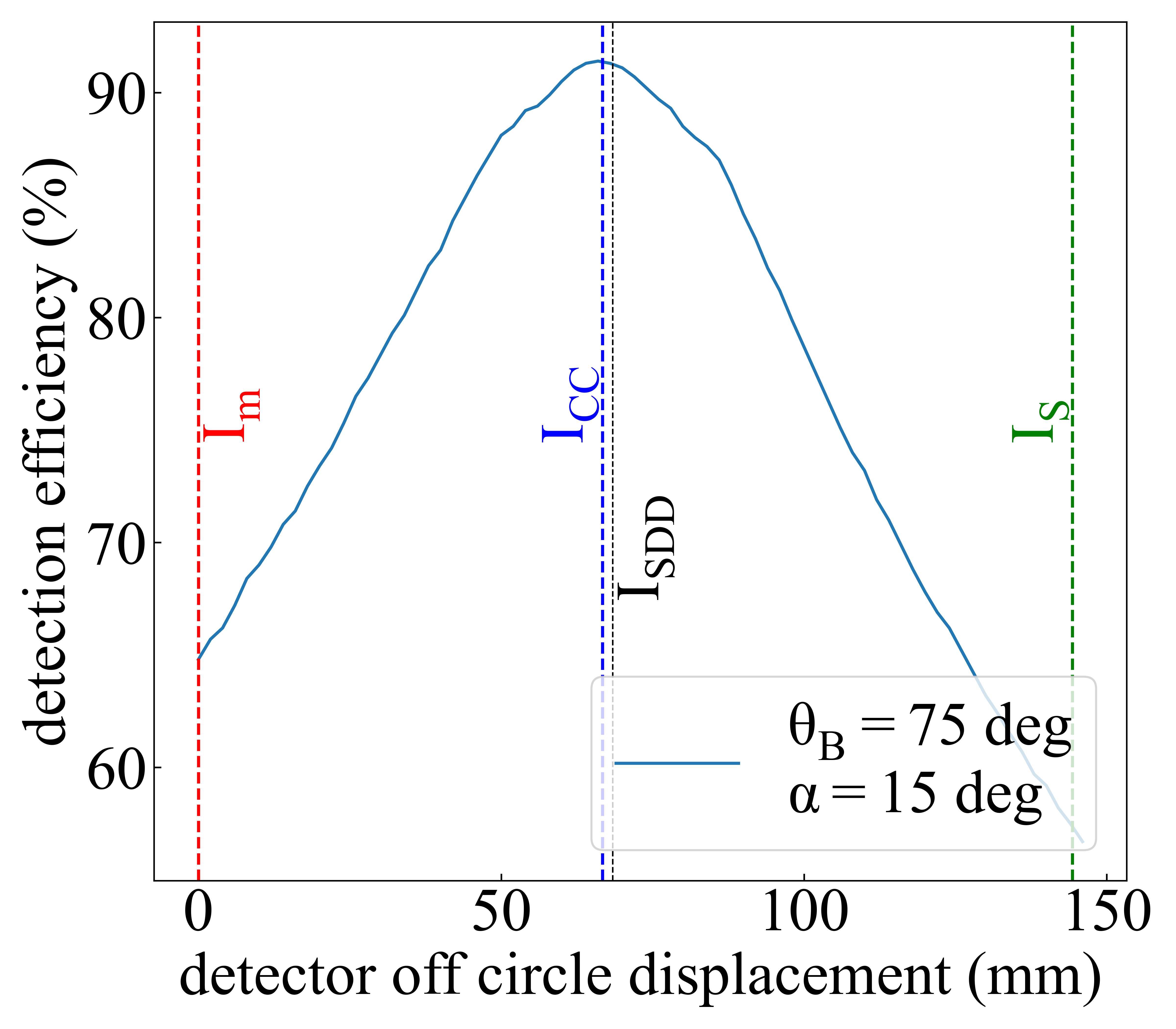}
  \includegraphics[width=\textwidth/2-4cm]{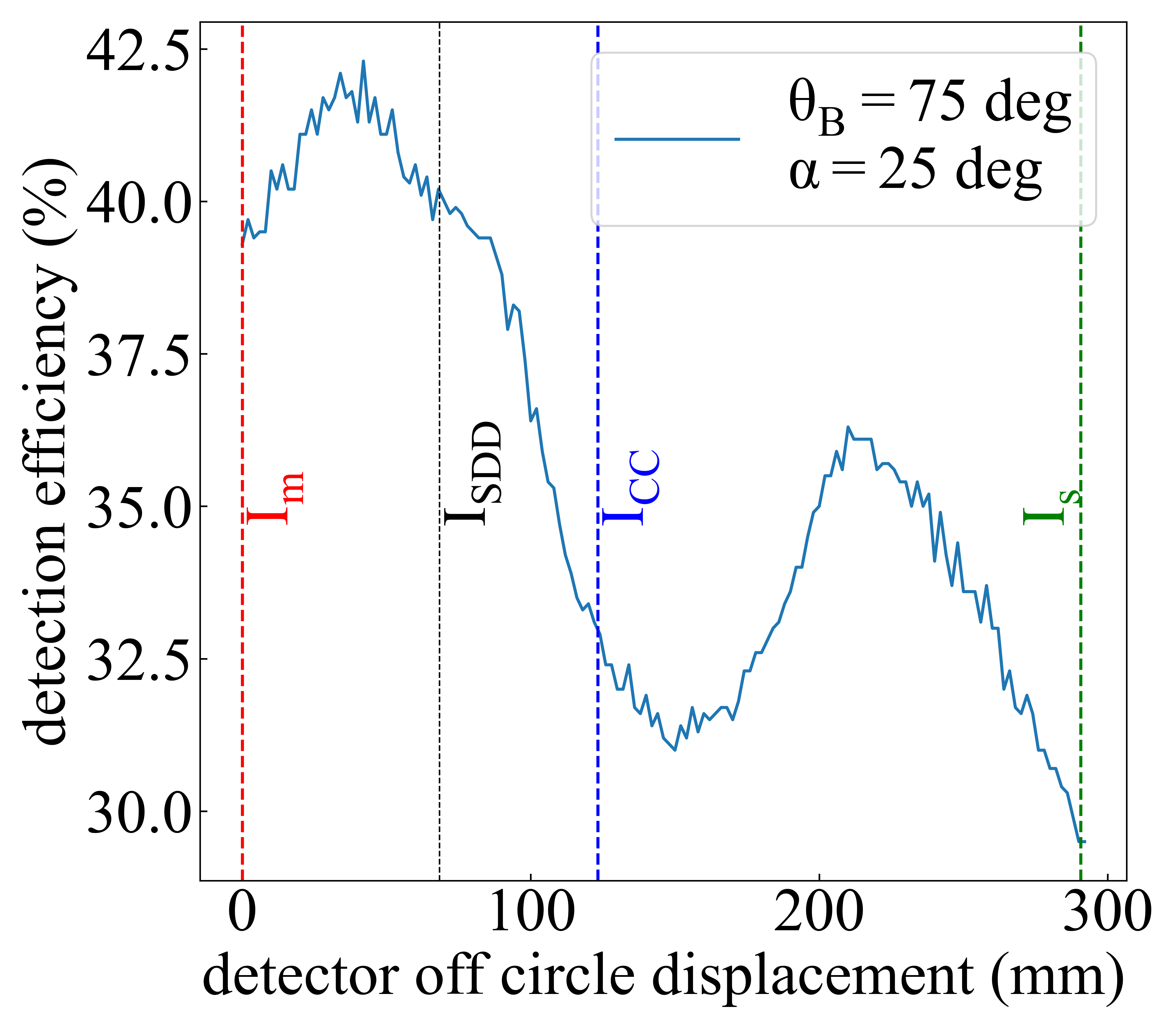}
  \caption{Efficiency of detector with diameter $D_{SDD}=13.8 \, \text{mm}$ as a function of the detector off-circle position, evaluated under both symmetric and asymmetric configuration at $\theta_B = 75 \, \text{deg}$.}
  \label{study4_det_pos_scan_75}
\end{figure}

To find the optimal detector location, we examine the detection efficiency (with detector diameter equals $D_{SDD}$) under symmetric and asymmetric configurations when $\theta_B = 75 \, \text{deg}$ as a function detector location (Figure \ref{study4_det_pos_scan_75}). The SBCA bent radius is set to be 500 mm and with a crystal face diameter of 100 mm. The position of $I_m$ (red), $I_s$ (green) and $I_{cc}$ (blue) is marked with dashed lines. An additional black line ($I_{SDD}$) at 68.4 mm marks the location where the beam horizontal width is equal to the circular detector sensor diameter.

We find the maximum efficiencies typically are at detector positions near $I_{cc}$, as suggested earlier, and often occurs at a location slightly in front of $I_{SDD}$ in all cases. Since the location of $I_{SDD}$ does not depend on $\theta_B$ and $\alpha$, the results suggest $I_{SDD}$ as a pragmatic static location for detector placement when utilizing asymmetric configuration.

\begin{figure}[!h]
\centering
  \includegraphics[width=\textwidth/2 -2cm]{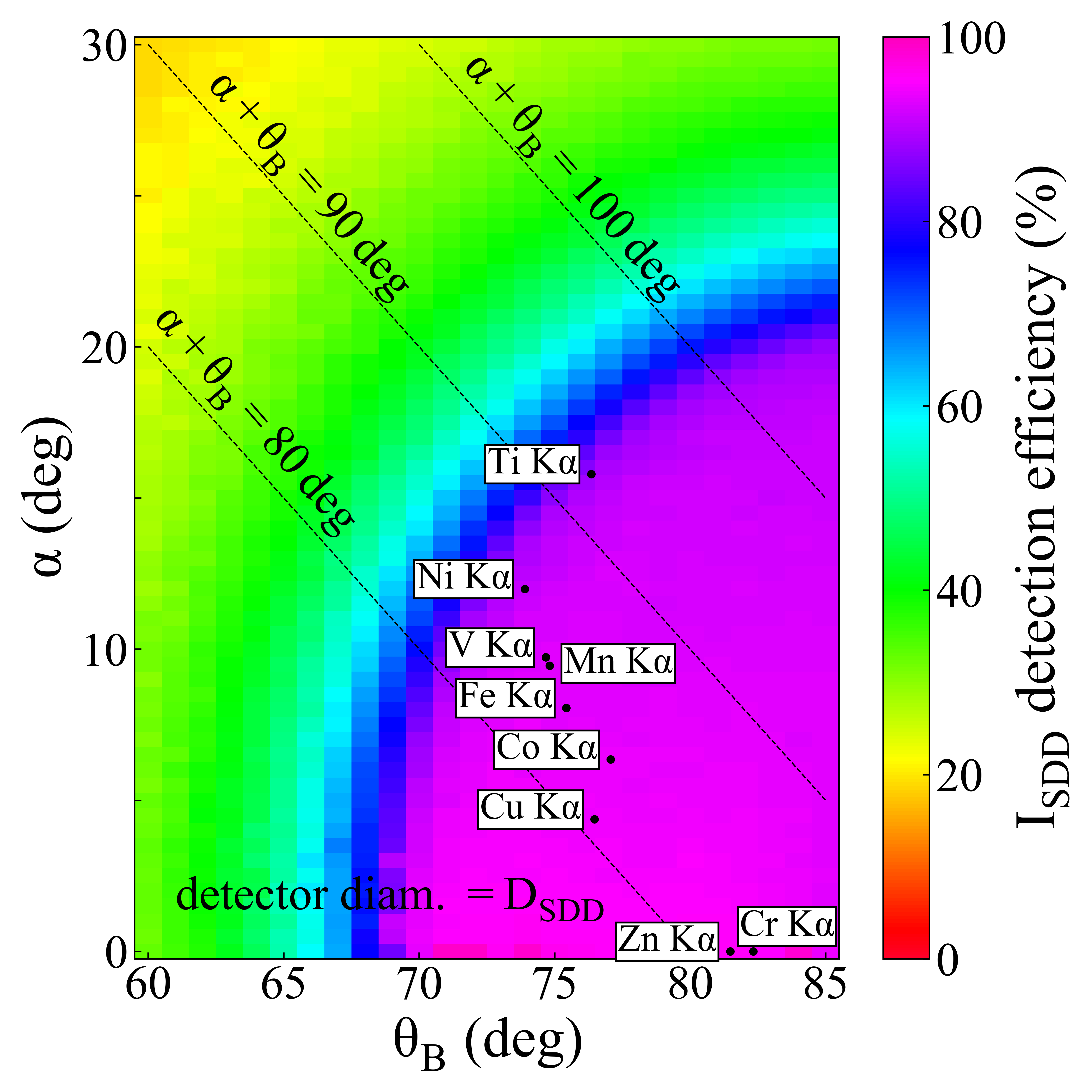}
  \caption{The detection efficiency for a detector with an active area diameter $D_{SDD}$ positioned at $I_{SDD}$. Labeled points indicate the best analyzer selection for the \textit{3d} transition metal K$\alpha_1$ emission lines, where $\theta_B$, $\alpha$, $G_0$, and $G_{hkl}$ for each emission line is detailed in Table \ref{reflection}. Here $R=500 \, \text{mm}$ and $d_{xtal}= 100 \, \text{mm}$.}
  \label{study4_det_eff_contour}
\end{figure}

To give a more holistic view, we quantitatively examine the effect of detector position on detection efficiency under asymmetric operation when the detector active area diameter equals $D_{SDD}$. Again, the issue here is the possible mismatch between the SDD active region size and the often larger dimension of the analyzed beam at the detector face. In Figure \ref{study4_det_eff_contour}, we show the detection efficiency at $I_{SDD}$ in the 2-D space of $\alpha$ and $\theta_B$, in which we observe that the detection efficiency exhibits an oval contour. We have also shown labeled points for the best analyzer selection (Si or Ge) for the \textit{3d} transition metal K$\alpha_1$ emission lines (with Scandium omitted due to its infrequent appearance in XES studies). Table \ref{reflection} gives the emission lines and associated analyzers, Bragg angles asymmetries, and reflections. While a full treatment of analyzer selection is outside of the scope of the present paper and is discussed elsewhere \cite{abramson24}, we have included the `best' performing reflection for context: typical losses are less than 30\% for lower-energy emission lines, while the plethora of possible \textit{d}-spacings as energy increases makes asymmetric operation lossless in terms of detection efficiency effects.

\begin{table*}[!]
\small
  \caption{Associated analyzers, bragg angles symmetries, and reflections for the labeled points in Figure \ref{study4_det_eff_contour}.}
  \label{reflection}
  \begin{tabular*}{\textwidth}{@{\extracolsep{\fill}}cccccc}
    \hline
    Emission & Energy (eV) & $\theta_B$ & $\alpha$ & $G_{hkl}$& $G_0$\\
    \hline
    Ti K$\alpha$  &   4510 & 76.35   & 15.79   & Ge(400) & Ge(511)   \\
    V K$\alpha$  &  4952 & 74.67 & 9.73 &   Ge(331) & Ge(553) \\
    Cr K$\alpha$  &  5416 & 82.33 & 0 & Ge(422) & Ge(422)   \\
    Mn K$\alpha$  &  5899 & 74.82 & 9.45 & Ge(511) & Ge(311) \\
    Fe K$\alpha$  &  6404 & 75.43 & 8.05 & Ge(440) & Ge(551) \\
    Co K$\alpha$  & 6930 &  77.07 & 6.35 & Si(531) & Si(642) \\
    Ni K$\alpha$  & 7478 &  73.9 & 11.99 & Ge(533) & Ge(642) \\
    Cu K$\alpha$  & 8048 & 76.47 & 4.37 & Ge(711) & Ge(511) \\
    Zn K$\alpha$  & 8639 & 81.48 & 0 & Si(642) & Si(642) \\
    \hline
  \end{tabular*}
\end{table*}

\subsection{Asymmetric Operation In X-ray Raman Imaging}

X-ray Raman scattering (XRS), sometimes also grouped under the term nonresonant inelastic x-ray scattering, uses hard x-rays to measure of the x-ray absorption spectrum for weakly bound shells \cite{tohji88, cramer98}.  This comes with three main advantages, all accruing from the use of high energy incident photons.  First, the higher penetrating power of hard x-rays ensures truly bulk-like measurement without risk of self-absorption effects. Second,  the high penetrating power, again, greatly simplifies and in some cases uniquely enables the measurement in special sample environments of the XAS spectrum for low-energy edges.  These uses have been most prominent in high pressure studies using diamond anvil cells \cite{mao11, sahle13}.  Finally, the large momentum transfers available from inelastic scattering of the incident photon allows tuning of selection rules, enabling measurement of a plethora of final state orbital angular momenta \cite{gordon08}.

However, these scientific benefits of XRS are inhibited by its very weak signal, requiring not just very high intensity synchrotron beamlines but also multianalyzer spectrometers seeking to maximize the collection solid angle. A complete discussion of such instrument design is outside the scope of this manuscript, but we note that the careful considerations of Huotari \textit{et al.}\cite{huotari17} illustrate well the reasons behind not just the continued use of 1-m radius of curvature optics but also the fact that even those optics are often masked to remove (or decrease) Johann error at some cost in collection solid angle.

\begin{figure}[!h]
\centering
  \includegraphics[width=\textwidth/2-5cm]{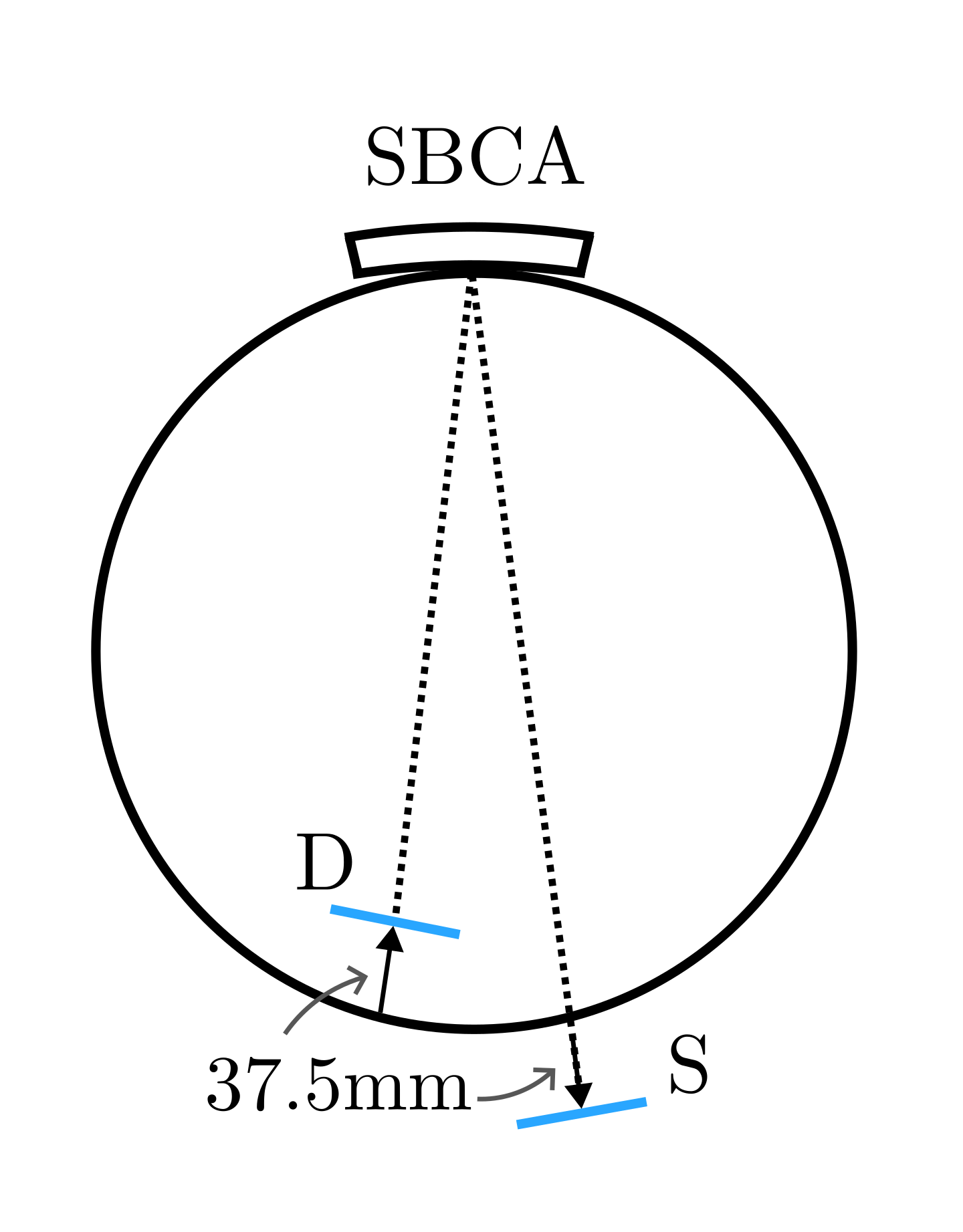}
  \includegraphics[width=\textwidth/2-2cm]{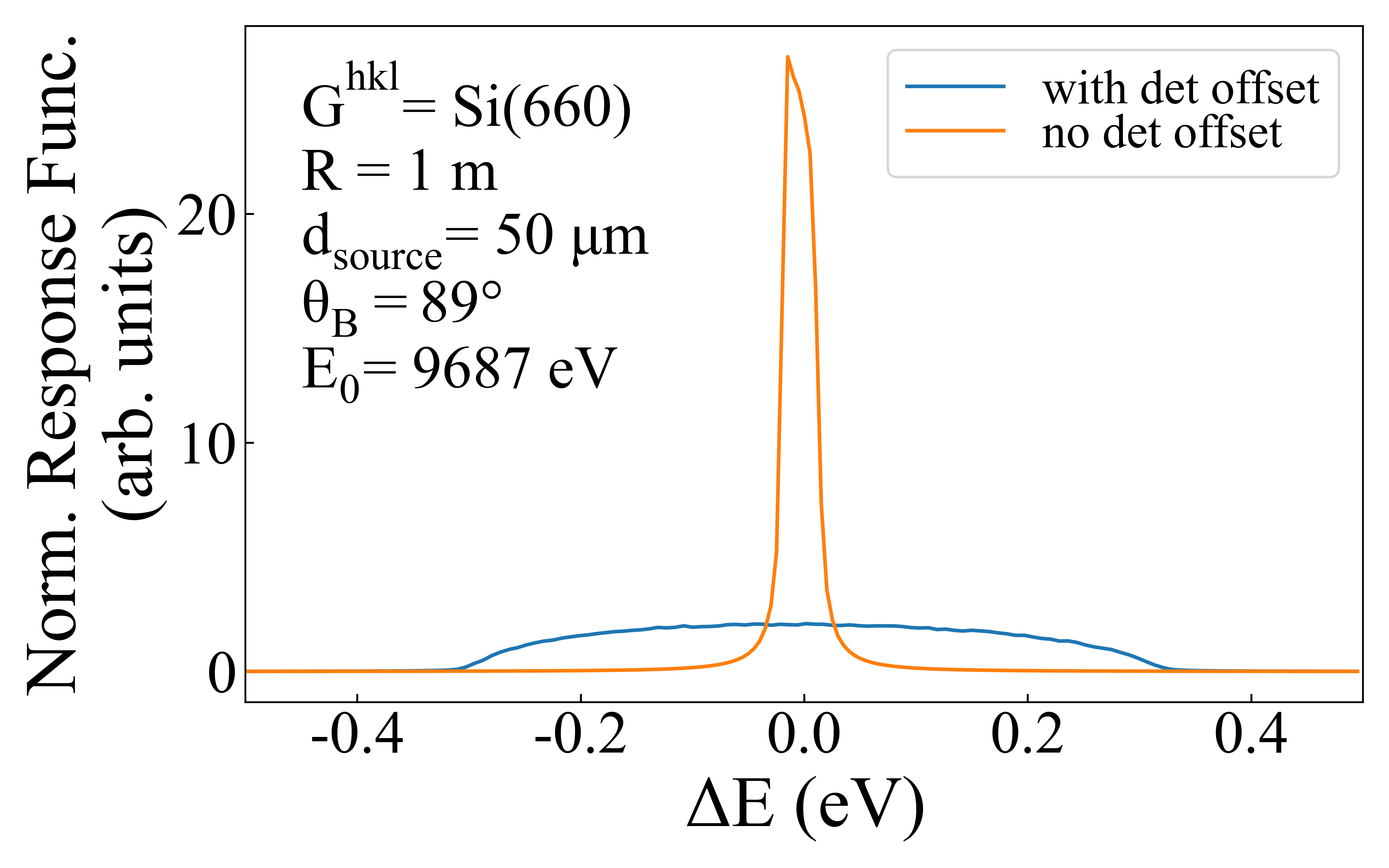}
  \caption{(Top) To gain more sample clearance and still maintaining focus when operating at higher $\theta_B$, the source and the detector is displaced off the Rowland circle symmetrically by 37.5 mm. (Bottom) Response functions at $\theta_B=89 \,\text{deg}$ when $R=1 \, \text{m}$ and: (i) the source and the detector is displaced off the Rowland circle by 37.5 mm (blue); (ii) no detector and analyzer offsets is applied (orange).}
  \label{XRS_det_offset}
\end{figure}

One use of XRS is Raman imagine, or direct tomography \cite{huotari11}. It uses the fact that extended source gives an extended image, to generate images with both spectral and spatial information. In the work of Huotari \textit{et al.}\cite{huotari17}, they reported that the state-of-the-art Raman imaging is constrained to near back scattering Bragg angles due to the energy broadening caused by Johann error and analyzer strains. To obtain more sample clearance but maintain quality of focus on the detector plane, the source and the detector are displaced off the Rowland circle which introduces another source of energy broadening (see Figure \ref{XRS_det_offset}). At $\theta_B = 89 \, \text{deg}$ with Si(660) 1-m radius SBCA, our ray tracing shows the energy broadening due a 75 mm detector offset is 0.52 eV. This is much larger than source sizes for any typical focused beam, and is a large part of the error budget.

Together with the maximization of the collection solid angle, these issues motivated Gironda \textit{et al.}\cite{gironda24}, to make use of the JNA configuration with the much larger, 0.5-m radius of curvature SBCA.  There, they make a first claim that the full solid angle of the newer optics (four times that of the traditional 1-m radius optics, even before masking) can be effectively used in XRS. Moreover, by utilizing the JNA configuration, operation at lower $\theta_B$ is allowed hence more sample clearance is obtained and no detector offset is required. This asserts a new paradigm for the design of such spectrometers: a small cluster of the larger, more tightly curved optics in JNA configurations will give a mechanically simpler and more cost-effective alternative to traditional designs while also increasing spectrometer solid angle and exhibits good energy resolution.

For bulk measurements, i.e., homogeneous sample not in special samples environments, the argument of Gironda \textit{et al.}\cite{gironda24}, is persuasive.  However, as admitted by those authors, a significant portion of XRS studies are performed in special sample environments where the XRS imaging methods that either reject the background inelastic x-ray scattering from sample chamber windows \cite{huotari17} or even complete spatial mapping of the XRS spectrum across a chemically inhomogeneous sample \cite{huotari11}.  We have previously mentioned (see Section \ref{det eff} and Figure \ref{study_4_det_imgs}) the larger aberrations of the analyzed radiation on the nominal detector plane when working asymmetrically on the Rowland circle.  Here, we address the consequences of those aberrations on XRS imaging applications.

\begin{figure}[!h]
\centering
  \includegraphics[width=\textwidth/2-4cm]{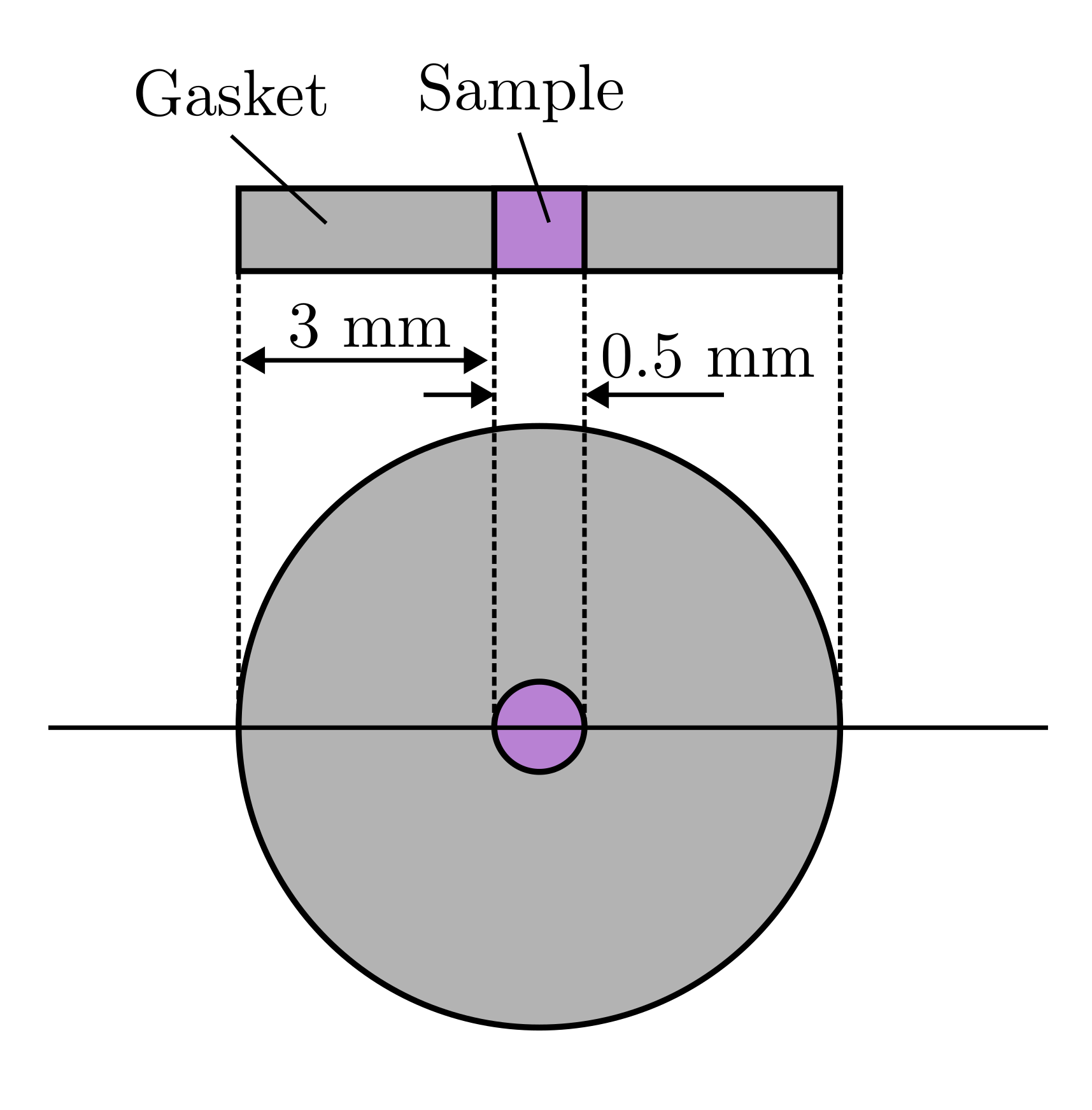}
  \caption{A schematic diagram for a typical diamond anvil cell used in x-ray Raman imaging where the top shows a sectional view seen a the Rowland plan, with a top down view shown at the bottom. The sample with 0.5 mm diameter is placed at the center of a gasket.}
  \label{diamond_anvil_cell}
\end{figure}

\begin{figure}[!h]
\centering
  \includegraphics[width=\textwidth/2-5cm]{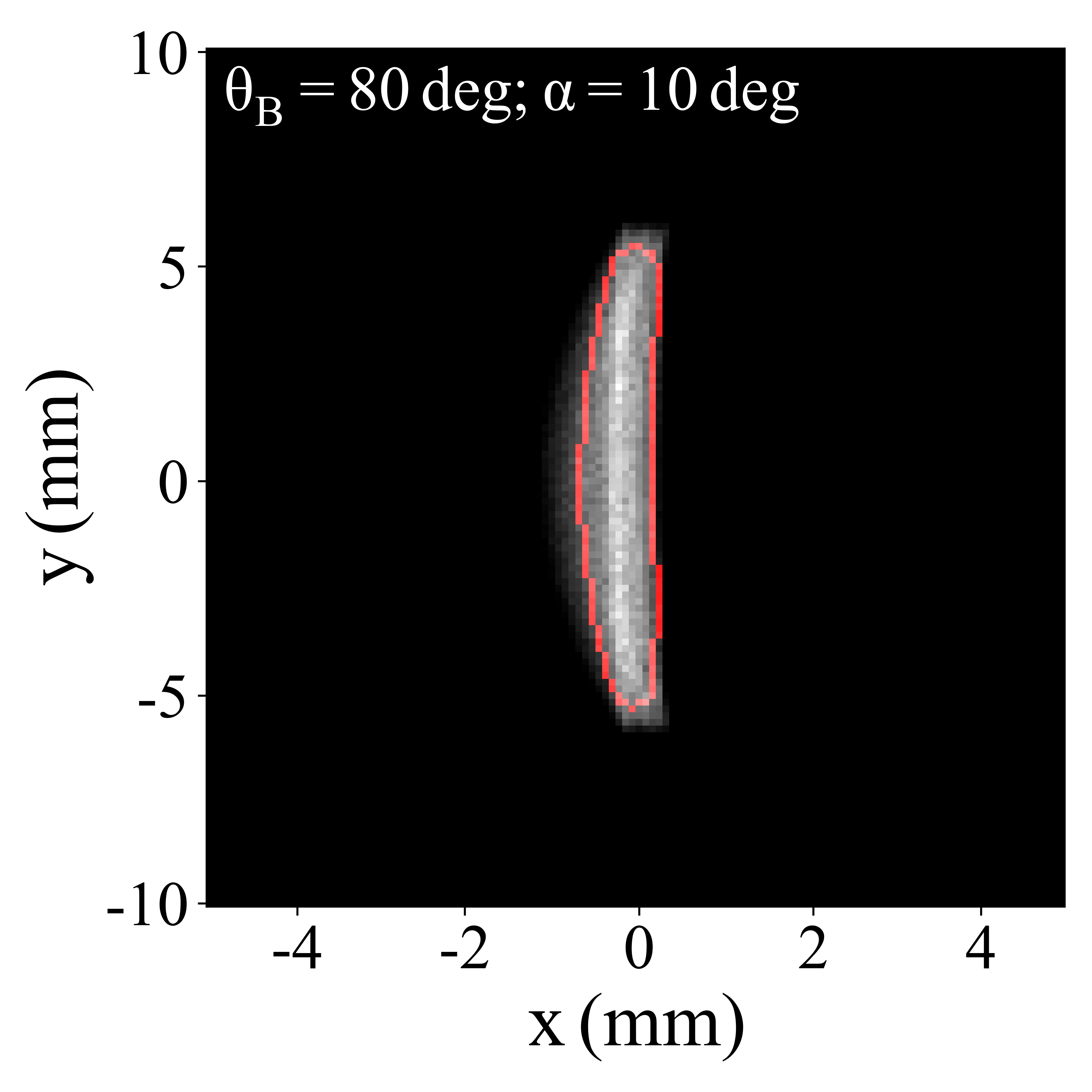}
  \includegraphics[width=\textwidth/2-5cm]{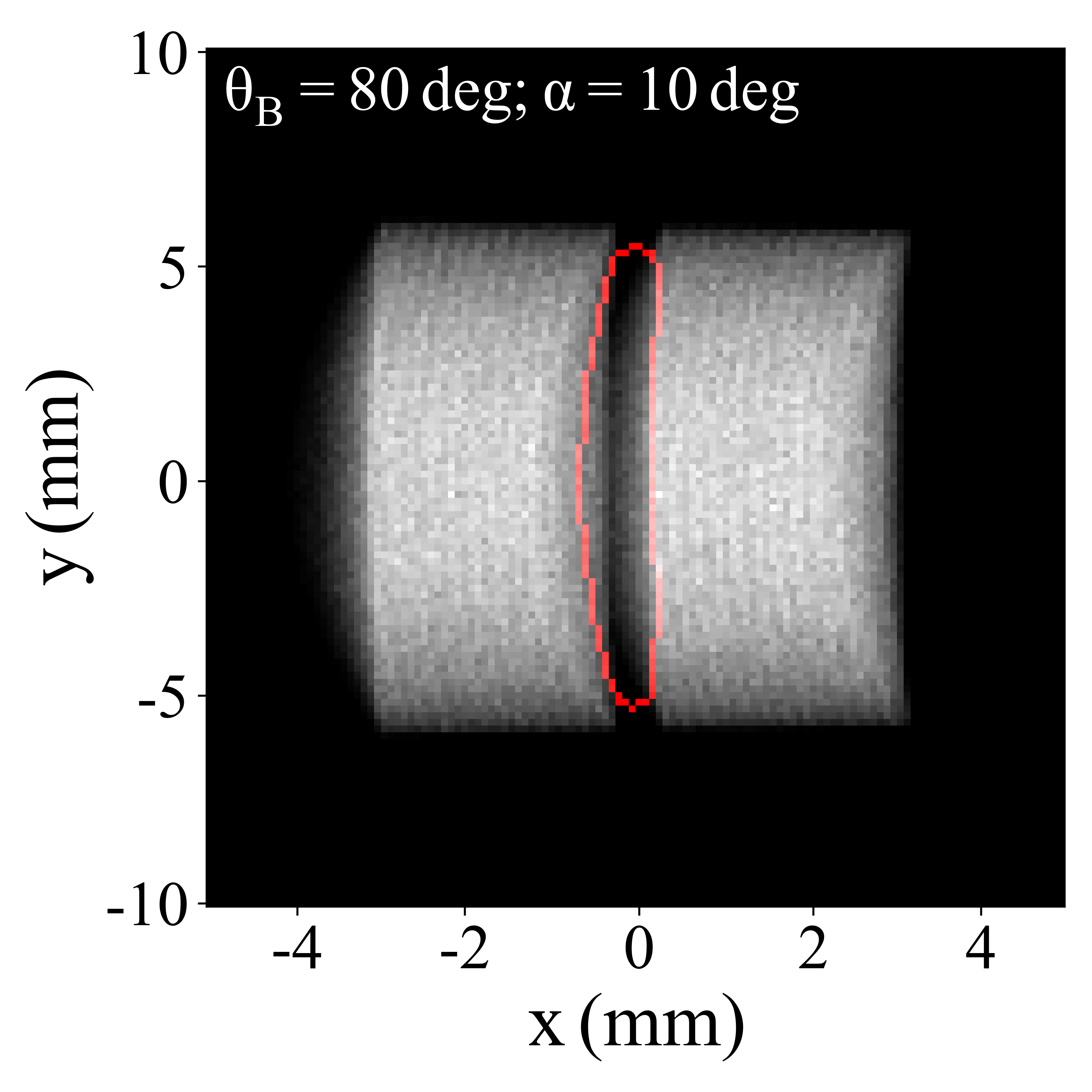}
  \caption{Detector image from the sample (left) and the gasket (right) under JNA when operating at $\theta_B= 80 \, \text{deg}$. The region outlined with the red indicates the ROI when calculating signal intensity. Note the the texture in the images is solely artifacts from plotting due to low image resolution.}
  \label{80_15_DAC}
\end{figure}

This is illustrated by means of a case study of a typical diamond anvil cell used in x-ray Raman imaging, as shown in Figure \ref{diamond_anvil_cell} \cite{sahle13}. We simulate this extended source under JNA at $\theta_B = 80 \, \text{deg}$ with multiple 50 $\mu$m sources spaced across the beampath through the gasket and sample. As shown in Figure \ref{80_15_DAC}, the detector-plane image produced by the sample (left panel) and the gasket (right panel) are examined separately. The red line defines the boundary of the suggested range of interest (ROI). It is taken to be the region where the pixel intensity is greater than 45$\%$ of the maximum intensity from the sample, and it captures roughly 75$\%$ of the sample signal.

To quantify the contamination of sample signal by the photons scattered from the gasket, we first calculate the percentage of the sample signal (57$\%$) and percentage of the environment signal (43$\%$) with respect to the total intensity (all within the given ROI), then the sample to environment signal ratio is found to be 1.3. This shows that asymmetric operation at $\theta_B = 80 \, \text{deg}$ (with $R=50 \, \text{cm}$) still maintains a high signal-to-environment ratio even with the detector image blurring, while the sample clearance increases from 8.2 cm (when using symmetric geometry with source and detector offset) to about 17 cm.

Although asymmetric operation offers advantages for XRS, it also presents limitations. When the sample environment shares the same element as is being studied in the sample, their signal peaks coincide in the Raman spectra. In such instances, any contamination from the environment is likely prohibitive. Similar concerns arise when using XRS to spatially map an extended, inhomogeneous sample -- this is best done with existing, symmetric SBCA configuration \cite{huotari17}.

\section*{Conclusions}
In this study, we present a comprehensive ray tracing analysis of the impact of asymmetric operation on a SBCA. We find a significant reduction in Johann error when operating asymmetrically, with complete elimination observed under the Johann Normal Alignment (JNA). We provide detailed insights into the standard deviation of the response functions ($\sigma_{resp}$) across a wide range of $\theta_B$ and $\alpha$, offering guidelines on best practice for asymmetric operation. Moreover, we explore the optimization of detector placement to mitigate signal losses resulting from sagittal image broadening under asymmetric operation. We propose a static location, $I_{SDD}$, for detector placement, and show that the detection efficiency remains close to maximum for a given $\theta_B$ and $\alpha$ under this arrangement. Finally, we extend our analysis to the effect of sagittal defocusing on X-ray Raman imaging, emphasizing the importance of spatial resolution, particularly in applications requiring sample environment signal rejection. Through the examination of a typical diamond anvil cell under JNA at $\theta_B = 80 \, \text{deg}$, we demonstrate that the signal-to-background ratio remains useful at 1.3, while increasing the sample clearance to about 17 cm (with half meter Rowland radius), highlighting the effectiveness of asymmetric operation in maintaining signal integrity and gaining sample clearance. 

In conclusion, our findings highlight the underutilization of asymmetric operation despite its manifold benefits. We foresee a widespread adoption of this approach in both synchrotron and laboratory settings in the near future. Moving forward, further research should extend beyond purely geometric calculations. Moreover, guidelines on the selection of analyzers and reflections that suits typical experimental constraints are needed \cite{abramson24}. With continued investigation, we can unlock the full potential of asymmetric operation, enhancing the capabilities of X-ray spectroscopy and imaging methods.

\section*{Conflicts of interest}
There are no conflicts to declare.

\section*{Acknowledgements}
This work is supported by funding from the U.S. Department of Energy in the Nuclear Energy University Program under Contract No. DE-NE0009158.



\balance


\bibliography{rsc} 
\bibliographystyle{rsc} 
\end{document}